\documentclass{emulateapj}

\newcommand{\beq}{\begin{equation}}
\newcommand{\beqa}{\begin{eqnarray}}
\newcommand{\eeq}{\end{equation}}
\newcommand{\eeqa}{\end{eqnarray}}

\newcommand{\bfx}{\mathbf{x}}

\newcommand{\bfeta}{\mbox{\boldmath{$\eta$}}} 

\shorttitle{Probability Distributions of Lensing Convergence, Shear,
  and Magnification}
\shortauthors{Takahashi et al.}

\begin{document}

\title{Probability Distribution Functions of Cosmological Lensing:
  Convergence, Shear, and Magnification}

\author{Ryuichi Takahashi\altaffilmark{1},
Masamune Oguri\altaffilmark{2,3},
Masanori Sato\altaffilmark{4},
Takashi Hamana\altaffilmark{3}}

\affil{\altaffilmark{1} Faculty of Science and Technology, Hirosaki
 University, 3 bunkyo-cho, Hirosaki, Aomori, 036-8561, Japan}
\affil{\altaffilmark{2} Institute for the Physics and Mathematics of the
 Universe, The University of Tokyo, 5-1-5 Kashiwa-no-ha, Kashiwa, 
 Chiba 277-8568, Japan}
\affil{\altaffilmark{3} Division of Theoretical Astronomy,
 National Astronomical Observatory of Japan,
 2-21-1 Osawa, Mitaka, Tokyo, 181-8588, Japan}
\affil{\altaffilmark{4} Department of Physics,
 Nagoya University, Chikusa, Nagoya 464-8602, Japan}

\begin{abstract}
We perform high resolution ray-tracing simulations to investigate
probability distribution functions (PDFs) of lensing convergence,
shear, and magnification on distant sources up to the redshift of
$z_s=20$. We pay particular attention to the shot noise effect in
$N$-body simulations by explicitly showing how it affects the 
variance of the convergence. We show that the convergence and
magnification PDFs are closely related with each other via the
approximate relation $\mu=(1-\kappa)^{-2}$, which can reproduce the
behavior of PDFs surprisingly well up to the high magnification tail. 
The mean convergence measured in the source plane is found to be
 systematically negative, rather
than zero as often assumed, and is correlated with the convergence
variance. We provide simple analytical formulae for the PDFs, which
reproduce simulated PDFs reasonably well for a wide range of redshifts 
and smoothing sizes. As explicit applications of our ray-tracing
simulations, we examine the strong lensing probability and the
magnification effects on the luminosity functions of distant galaxies
and quasars. 
\end{abstract}


\section{Introduction}

When light rays from distant sources propagate through the
inhomogeneous matter distribution in the Universe, they are scattered
many times by intervening clumps of matter.  
Because of this gravitational lensing effect (e.g. Schneider, Ehlers
\& Falco 1992;  Schneider, Kochanek \& Wambsganss 2006),
a light bundle that propagates through the overdense (underdense)
region is magnified (demagnified) and is also deformed.
Therefore, observed brightnesses, sizes, and shapes of distant 
objects can in fact differ from those computed assuming the
homogeneous universe, suggesting that all cosmological observations
are subject to gravitational lensing. 

Gravitational lensing effects are more pronounced for more distant
sources. Thus it is of growing important in modern astronomy where
many high redshift objects are being discovered.
For instance, Bouwens et al. (2010a,b) reported candidates of galaxies
at $z \gtrsim 7$ in deep images taken with the Hubble Space
Telescope, and derived the luminosity function at the redshift (see
also Oesch et al. 2010; Yan et al. 2010). 
Lehnert et al. (2010) presented the spectroscopy of a distant
Lyman-$\alpha$ emitting galaxy, whose redshift of $z=8.6$ turned out
to be higher than those of previously known most distant objects,
including a gamma-ray burst at $z=8.2$ (Salvaterra et al. 2009; Tanvir
et al. 2009) and a galaxy at $z=6.96$ (Iye et al. 2006). The recent
progress in the exploration of the deep universe made it possible to
derive accurate luminosity functions of $z$-dropout galaxies and 
Lyman-$\alpha$ emitters at $z \sim 7$ (Ouchi et al. 2009, 2010; Ota et
al. 2010). It is clear that one has to take proper account of
gravitational lensing effects for studying intrinsic properties of
these very distant sources (see, e.g., Wyithe et al. 2011, for a
recent study of the lensing effects on very high redshift galaxies at
$z \gtrsim 10$). 

Turning the problem around, we can even make use of lensing
magnification as a `cosmic telescope' to detect and explore faint
high-redshift objects that are not accessible without the help of lensing.
Indeed, several highly magnified galaxy candidates at up to $z \sim
10$ have been discovered behind massive clusters (Bouwens et al. 2009;
 Bradley et al. 2008; Richard et al. 2006, 2008; Stark et al. 2007;
 Bayliss et al. 2010). Lensing magnification can provide a way to
 locate very high-redshift supernovae as well (e.g., Oguri, Suto \&
 Turner 2003; Goobar et al. 2009; Oguri \& Marshall 2010). 
 Upcoming wide-field surveys such as Hyper Suprime-Cam (HSC; Miyazaki et
 al. 2006), the Dark Energy Survey (DES)\footnote{home page: http://www.darkenergysurvey.org/} and Large Synoptic Survey Telescope (LSST; LSST Science
 Collaborations et al. 2009) will find more such highly magnified
 distant objects, which serve as good follow-up targets for the next
 generation telescopes such as the Thirty Meter Telescope (TMT)\footnote{home
   page: http://www.tmt.org/} and the James Webb Space Telescope
 (JWST)\footnote{home page: http://www.jwst.nasa.gov/}.

Recently lensing magnification attracts much attention in the survey of
submillimeter galaxies. Because of the steep number counts of
submillimeter galaxies, it is expected that their observed number
counts are significantly modified particularly at the bright end such
that bright submillimeter sources are almost exclusively highly
magnified lensing events. This prediction has been confirmed by
Negrello et al. (2010), who showed that bright submillimeter sources
from the Herschel Astrophysical Terahertz Large Area Survey are indeed
dominated by strong lensing events. Similar enhancements of the number
density at the bright end was pointed out by van der Burg et
al. (2010) for the UV galaxy luminosity function at $z=3-5$ from the
Canada-France-Hawaii-Telescope Legacy Survey (CFHTLS). 

Furthermore, measurements of distances to any distance indicators,
which serve as one of the most fundamental ways to constrain
cosmological parameters including dark energy, are always affected by
gravitational lensing.  For instance, gravitational lensing not only
induces an external dispersion in distance-redshift relations
derived from standard candles/sirens such as type Ia supernovae, gamma
ray bursts, and the gravitational waves from neutron star binaries and
 binary black holes (e.g., Hamana \& Futamase 2000; Hirata et
 al. 2010; Shang \& Haiman 2011), 
but also may cause systematic
biases in the derived relations given the non-Gaussian nature of the
probability distribution of lensing magnification (e.g., Sarkar et al. 2008). 
Kronborg et al. (2010)  found an evidence of lensing in  the distant type-Ia
supernovae with a $2 \sigma$ significance level, from the analysis of
the Supernova Legacy Survey data. The external convergence caused by
lensing is also known as one of the most significant sources of
systematic effects in constraining the Hubble constant from time
delays between quasar images (e.g., Oguri 2007; Suyu et al. 2010)

In order to address the effects of gravitational lensing mentioned
above, we need an accurate and reliable model of the probability
distribution function (PDF) of gravitational lensing. 
A powerful way to predict the PDF in a given cosmological model is to
resort to the ray-tracing simulations. In fact ray-tracing simulations
have been used to study a wide range of gravitational lensing effects,
including the magnification PDF $P(\mu)$ (Refsdal 1970;
Schneider \& Weiss 1988a,b; Jaroszynski et al. 1990; Wambsganss et al. 1998;
Tomita, Asada \& Hamana 1999; Tomita, Premadi \& Nakamura 1999;
Barber et al. 2000; Hamana et al. 2000; Wang et al. 2002; Takada \&
Hamana 2003; Yoo et al. 2008), the convergence PDF $P(\kappa)$ (Jain
et al. 2000; Taruya et al. 2002; Das \& Ostriker 2006), the shear PDF
$P(\gamma)$ (Barber et al. 2000; Jain et al. 2000),  
 the relations among the magnification, convergence and shear
 (Barber et al. 2000; Hilbert et al. 2011),
 the strong lensing probability (Wambsganss et al. 1995;
 Bartelmann et al. 1998; Meneghetti et al. 2005; Hilbert et al. 2007),
 two point angular correlation function of the magnification
 (Takada \& Hamana 2003), and the cosmic shear (Jain et al. 2000; Hamana
 et al. 2002; Vale \& White 2003; White \& Vale 2004; Semboloni et al. 2007;
 Hilbert et al. 2009; Sato et al. 2009, 2011; Sato, Ichiki \& Takeuchi 2011).
However, accuracies of many of previous ray-tracing simulations are
limited by the resolution. While we have to resolve the sub-galactic
scales ($< 10$~kpc) in order to make a reliable prediction for the
magnification of distant galaxies, only two simulations reached the
galactic-scale resolution. One is the simulation by Wambsganss et al.,
who used the small box $5h^{-1}$Mpc on a side with $256^3$ dark matter
particles, the mean particle separation is $20h^{-1}$kpc and the
softening length is $10h^{-1}$kpc (Cen et al. 1994; Wambsganss et
al. 1995, 1997, 1998). They calculated the strong lensing probability,
the distribution of the  magnification, and the lensing effects on the
determination of the  deceleration parameter $q_0$ using type-Ia
supernovae. 
The other one is the work of Hilbert et al., who used the Millennium
Simulation (Springel et al. 2005) in which the box is $500h^{-1}$Mpc
on a side with $2160^3$ particles, the mean particle separation is
$230h^{-1}$kpc and the softening length is $5h^{-1}$ kpc (Hilbert et
al. 2007, 2008, 2009). They discussed many of the issues such as
$P(\mu)$, $P(\kappa)$, the strong lensing probability and the effects
of baryon. 

An alternative method to predict lensing PDFs is to use the so-called
halo model (e.g., Jain \& Lima 2011; Lima et al. 2010a,b; Kainulainen
 \& Marra 2009,2011, for a recent
example). While the halo model can make accurate predictions of
PDFs at the high magnification tail where contributions from single
halos dominate, predicting the PDFs near the peaks with the halo model
is quite challenging. Thus, cross-checking of the halo model with
high-resolution ray-tracing simulations are crucial for validating and
possibly improving the accuracy of the halo model predictions. 

In this paper, we perform high resolution ray-tracing simulations to
study lensing PDFs in great detail. The box size of $50h^{-1}$Mpc with
$1024^3$ particles, the mean particle separation of $50h^{-1}$kpc, and
the softening length of $2h^{-1}$kpc indicates that our simulations
represent, to our knowledge, the highest resolution ray-tracing
simulation conducted to date for studying cosmological lensing effects.
We run three sets of simulations with $256^3$, $512^3$ and $1024^3$
particles to check the numerical convergence of our simulation
results, and also compare the two box sizes of $50h^{-1}$Mpc and
$100h^{-1}$Mpc to check the effects of density fluctuation larger than
the box size. We consider the gravitational evolution of the dark
matter particles only, and do not include the effect of baryon cooling
which would enhance the strong lensing probability (e.g., Hilbert et
al. 2008). In this paper, we present comprehensive analysis of PDFs of
convergence, shear, and magnification, with a particular emphasis on
the relation between these three lensing quantities. 
Our exploration of the PDF up to very high redshift of $z_s=20$
enables the immediate application of our results for studying
high-redshift ($z\gtrsim 7$) sources that are recently discovered and
also predicting even more distant sources that will be discovered
in the future.  
Our numerical results of two-dimensional maps of lensing fields and
PDFs are made publicly available at 
http://cosmo.phys.hirosaki-u.ac.jp/takahasi/raytracing/.

The structure of the present paper is as follows.
In Section~\ref{sec:sim}, we describe the details of $N$-body
simulations and our ray-tracing simulations and method to obtain
two-dimensional maps of the convergence, the shear and the
magnification.
In Section~\ref{sec:result}, we show the PDFs of the convergence, the shear and
the magnification and provide simple analytical formulae of these PDFs
which well reproduce PDFs obtained from ray-tracing simulations over a
wide range of redshifts and smoothing scales.
In Section~\ref{sec:stronglens}, we examine the strong lensing
probability by categorizing lensing mapping into tree types.
In Section~\ref{sec:mag}, we discuss magnification effects on luminosity
functions of distant sources.
Section~\ref{sec:summary} is devoted to summary and discussion.

Throughout the present paper, we adopt the standard $\Lambda$CDM model
with matter density $\Omega_{m} =0.274$, baryon density
$\Omega_{\rm b}=0.046$, dark energy density $\Omega_{\rm w}=0.726$,
spectral index $n_{\rm s}=0.96$, amplitude of fluctuations $\sigma_8=0.812$, 
and expansion rate at the present time $H_{0}=70.5$km s$^{-1}$ Mpc$^{-1}$,
consistent with the WMAP 5-year results (Komatsu et al. 2009).

\section{Numerical Simulations}
\label{sec:sim}
\begin{figure}
\vspace*{1.0cm}
\plotone{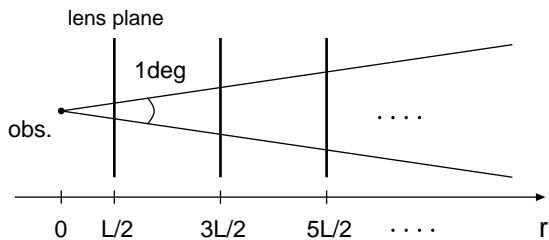}
\caption{
A schematic picture of our ray-tracing simulation.
The horizontal axis is the comoving distance $r$ from the observer.
The vertical thick lines denote the positions of the lens planes,
which are located at $r=L \times (i+1/2)$ with $i=0,1,2,\cdot \cdot$.
The source planes are located at $r=L \times i$.
Light rays are emitted from the observer and are scatted at the lens
planes before reaching the source plane. 
The field of view is $1 \times 1 {\rm deg}^2$.
}
\label{fig_lens_planes}
\vspace*{0.5cm}
\end{figure}

Figure~\ref{fig_lens_planes} shows a schematic picture of our ray-tracing
simulation. The horizontal axis is the comoving distance $r$ from the
observer, and thick vertical lines are the lens planes. The lens
planes are placed at an equal distance intervals of $L$,  $r=L \times
(i+1/2)$ with an integer $i=0,1,2,\cdot \cdot \cdot$. We place the
lens planes from the observer to the highest redshift of $z_s=20$ with
the interval $L=50h^{-1}$Mpc and $100h^{-1}$Mpc. Light rays are
emitted from the observer and are deflected at each lens plane before
reaching the source plane.  Source planes are placed in between the
lens planes, i.e., at $r=L \times i$.
In our simulation, the field of view is $1 \times 1 {\rm deg}^2$.
We impose the periodic boundary condition on the lens planes.

\subsection{$N$-body Simulations}

We run the $N$-body simulations on the cubic box, and then project the
particle positions to the two dimensions, in order to obtain the
particle distribution and the gravitational potential on the lens planes.
We use the numerical simulation code Gadget2 (Springel, Yoshida \&
 White 2001; Springel 2005).
We generate the initial conditions based on the second-order Lagrangian
 perturbation theory (2LPT; Crocce, Pueblas \& Scoccimarro 2006;
 Nishimichi et al. 2009) with the initial linear power spectrum calculated
 by Code for Anisotropies in the Microwave Background (CAMB; Lewis,
 Challinor \& Lasenby 2000)\footnote{see also http://camb.info/}.
We use a parallelised 2LPT code which is kindly provided by Takahiro
  Nishimichi (Valageas \& Nishimichi 2011) to run large cosmological
  $N$-body simulations with initial conditions based on 2LPT.
We dump the outputs (the particle positions) at the redshifts
corresponding to the positions of the lens planes $r=L \times (i+1/2)$,
shown in Figure~\ref{fig_lens_planes}.
The size of the simulation box is $L$ on a side. We have checked that
the matter power spectra of our $N$-body simulations agree with the
results of the higher resolution simulation, in which we used the
finer simulation parameters of the time step, the force calculation,
etc., within $2 (10)\%$ for $k < 20 (80) h/$Mpc.

Table~\ref{table1} lists four models of our simulations.
The table shows the simulation box on a side $L$, the number of
 particles $N_p^3$, the particle mass $m_p$, the softening length
 $r_s$, the initial redshift $z_{\rm in}$,
 and the number of lens planes $N_{\rm lens}$ up to $z_s=20$.
The softening length is fixed to be $4\%$ of the mean particle 
 separations for all the models. 
The first three models are $L=50h^{-1}$Mpc with different number of
 particles, $256^3$, $512^3$, and $1024^3$, which we name S256, S512, and
 S1024, respectively. We use these three models to check the numerical
 convergence of our simulation results at small scales, particularly
 because the lensing magnification is known to be very sensitive to
 the mass resolution. As we will show later, the poor mass resolution
 simulations are indeed affected significant by the shot noise.
 The model L512, $L=100h^{-1}$Mpc with $N_p^3=512^3$, is the same 
 resolution as S256 but the simulation volume is eight
 times larger than S256. We use this model to check the effect of
 density fluctuation larger than the box size (see Appendix A).
 For S256, S512 and L512, we prepare the different realizations at each
 lens plane to reduce the sample variance\footnote{We run the $N$-body
 simulation for the $i$-th lens plane at $r=L\times (i-1/2)$ only
 from the initial redshift to the redshift of the lens plane.}.
 However, for S1024, we perform only four independent realizations
 because of the limited computer resources, and we repeatedly use the
 outputs at different redshifts to construct a light-cone output.

\begin{deluxetable*}{lcccccc}
\startdata
  \hline
 Models  & $L$ ($h^{-1}$Mpc)  & $N_p^3$ & $m_p(h^{-1} M_\odot)$ &
 $r_s(h^{-1}{\rm kpc})$ & $z_{\rm in}$ & $N_{\rm lens}$ \\ \hline 
  S256 & $50$ & $256^3$  & $5.7 \times 10^8$ & $8$ & 80 & 158 \\
  S512 & $50$ & $512^3$  & $7.1 \times 10^7$ & $4$ & 90 & 158 \\
  S1024 & $50$ & $1024^3$ & $8.9 \times 10^6$ & $2$ & 100 & 158 \\
  L512 & $100$ & $512^3$  & $5.7 \times 10^8$ & $8$ & 80 & 79
\enddata
\tablecomments{
Models of our ray-shooting simulations. The cubic simulation box on
 a side $L$, the number of particles $N_p^3$, the particle mass $m_p$,
 the softening length $r_s$, the initial
 redshift $z_{\rm in}$, and the number of lens planes $N_{\rm lens}$
 up to $z_s=20$.
}
\label{table1}
\end{deluxetable*}

\subsection{Ray-tracing Simulations}
We briefly explain the procedure to trace light rays through $N$-body
 data and obtain the maps of the lensing fields on the source plane.
We use the code RAYTRIX (Hamana \& Mellier 2001) which follows the
 standard multiple lens plane algorithm.
In the standard multiple lens plane algorithm, the distance between
 observer and source galaxies is divided into several intervals.
In our case, as shown in Figure~\ref{fig_lens_planes}, we adopt a
fixed interval whose value is the same as simulation box $L$ on a
side. Particle positions are projected onto two dimensional lens planes
 ($xy$, $yz$ ,$zx$ planes) every $L$.
 Using Triangular-Shaped Cloud method (Hockney \& Eastwood 1988), 
 we assign the particles onto $N_g^2$ grids in lens planes, then compute
 the projected density contrast at each plane. We test the convergence
 of our simulation by varying resolution from $N_g^2=256^2$ to $16384^2$. 
 The two-dimensional gravitational potential is solved via Poisson
 equation using Fast Fourier Transform. Finally, two dimensional sky
 maps of the convergence, the shear, the magnification, and the
 angular positions of light rays are obtained by solving the evolution
 equation of Jacobian matrix along the light-ray path which is obtained
 by solving the multiple lens equation.

 We prepare $20$ realizations by randomly choosing the projecting
 direction and shifting the two dimensional positions.
 In each realization, we emit $2048^2$ light-rays, leading to $8
 \times 10^7$ rays in total. 

We note that our work is complementary to the previous work by Sato et
al. (2009,
 2011) who conducted ray-tracing simulations using the similar
 technique as described above. Their interests lie in accurate
 predictions for the covariance matrices of the cosmic shear power
 spectrum and correlation function, which are required for extracting
 cosmological information from future wide field optical imaging surveys
 (e.g., HSC survey, DES and LSST), 
whereas our purpose is to predict the lensing PDFs for
 distant sources, for which high resolution simulations are crucial.

\section{Results}
\label{sec:result}
\subsection{Variance of the Convergence}
\label{sec_kappa}

The surface area of the ray bundle in the source plane is smaller
(larger) than that in the image plane due to the cosmic magnification
(demagnification). As a result, the probability distributions of the
convergence, shear and  magnification evaluated in the image plane
differs from those in the source plane. Throughout this paper, we will
show the PDFs defined in the source plane, because our interest lies
in the predictions of lensing effects on the distant sources, for
which source-plane lensing PDFs are more relevant. Since in our
ray-tracing simulation we emit the light rays homogeneously from the
observer to the first lens plane, we can easily derive the
source plane PDFs by adding a weight of the magnification in computing
the PDF from simulations (see below).

In this section, we show our numerical results of the variance of the
 convergence $\kappa$ in the source plane :
\beq
 \langle \kappa^2 \rangle = \frac{\sum_j \left( \kappa_{j}^2/\mu_{j} \right)}
                           {\sum_j  \left( 1/\mu_{j} \right)}.
\eeq
Here $\kappa_j$ and $\mu_j$ denote the convergence and magnification 
for $j$-th light ray, respectively. The summation runs over all the
light rays of $8 \times 10^7$. The factor $1/\mu_j$ originates from
the ratio of the area in the image and source planes.

\begin{figure}
\epsscale{1.0}
\plotone{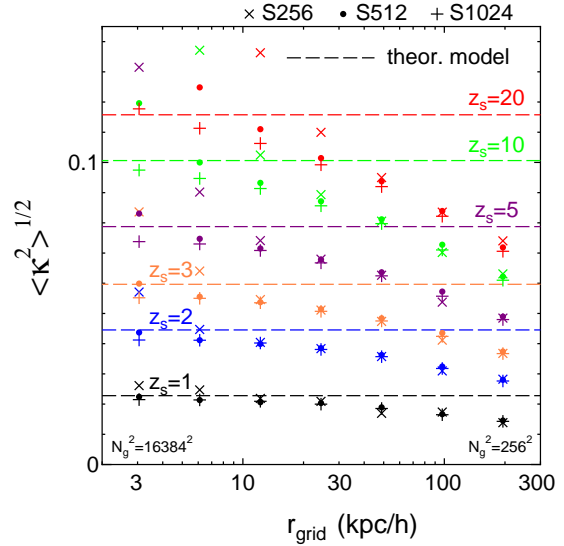}
\caption{
Root-mean-square of the convergence, $\langle \kappa^2 \rangle^{1/2}$,
 as a function of the smoothing scales $r_{\rm grid}$ at the source
 redshifts of $z_s=1,2,3,5,10$ and $20$.
Different symbols denote simulation results with different mass
 resolutions, the cross $\times$ for $N_p^3=256^3$, the circle
 $\bullet$ for $512^3$ and the plus $+$ for $1024^3$, respectively. 
 The dashed lines are the weak lensing prediction given in
 Equation~(\ref{rms_kappa}). 
}
\label{fig_kappa}
\vspace*{0.5cm}
\end{figure}

Figure~\ref{fig_kappa} shows the root-mean-square of the convergence,
 $\langle \kappa^2 \rangle^{1/2}$.
The horizontal axis is the grid size of the two dimensional gravitational
 potential in the lens planes.
Here the grid size is $r_{\rm grid}=L/N_g$ with
 the number of grid $N_g^2$ of the gravitational potential, and
 our finest resolution is
 $r_{\rm grid}=3h^{-1}$kpc (corresponding to $N_g^2=16384^2$).
The symbols are our simulation results, the cross $\times$, the circle
 $\bullet$ and the plus $+$, correspond to the various number of particles,
 $N_p^3=256^3$ (model S256), $512^3$ (S512) and $1024^3$ (S1024),
 respectively.
The source redshifts are $z_s=1,2,3,5,10$ and $20$.

 In the weak lensing approximation, the variance of the convergence is
 given by (e.g., Bartelmann \& Schneider 2001),
\beqa
 \langle \kappa^2 \rangle = \frac{9}{8 \pi} H_0^4 \Omega_m^2 \int_0^{z_s}
 \frac{dz}{H(z)} \left( 1+z \right)^2 \left[ \frac{r(z) r(z,z_s)}{r(z_s)}
 \right]^2  \nonumber \\
 \times \int dk k P(k,z),
\label{rms_kappa}  
\eeqa
where $r(z,z_s)$ is the comoving distance from $z$ to $z_s$ and  
$P(k,z)$ is the matter power spectrum as a function of the wavenumber
 $k$ and $z$. We calculate the matter power spectrum $P(k,z)$ in
 Equation~(\ref{rms_kappa}) directly using our $N$-body simulation
 data at the redshifts of the lens planes\footnote{To calculate the density 
   fluctuations in the cubic  box, we assign the particles onto a
   $1280^3$ grid using the cloud-in-cell method. Then we perform the
   Fourier transform to calculate the power spectrum.}. The
 predictions are shown by horizontal dashed lines in Figure~\ref{fig_kappa}.
We do not use the theoretical fitting formula of the halo-fit model
 (Smith et al. 2003) to calculate the non-linear power spectrum,
 because it has recently been reported that the halo fit
 underestimates the power spectrum  at small scales, $k>0.1h/$Mpc
 (e.g. White \& Vale 2004; Sato et al. 2009; Heitmann et al. 2010).
 For reference, the variance in Equation~(\ref{rms_kappa}) becomes smaller
 by a few ten percents if we use the halo-fit model to compute the
 non-linear matter power spectrum $P(k,z)$ in Equation~(\ref{rms_kappa}).

 As shown in Figure~\ref{fig_kappa}, the simulation results 
 decrease for larger $r_{\rm grid}$. This is because the density
 fluctuations smaller than $r_{\rm grid}$ are smeared out when
 assigning the particles onto the grids. On the other hand, for
 smaller $r_{\rm grid}$, our simulation results are larger than the
 theoretical prediction (the dashed lines) especially for the higher
 redshifts and for the lower mass resolution simulations, which we
 ascribe to the shot noise. For higher source redshift, the smaller
 density fluctuations  generate the convergence $\kappa$ in
 Equation~(\ref{rms_kappa})  (the peak position of integrand in
 Equation~(\ref{rms_kappa}), $k^2 P(k,z)$, shifts larger $k$ for higher $z$).
Hence, in order to study the lensing of high redshift sources
 using the ray-tracing simulations, we need sufficient number of particles
 to reduce the shot noise.

\begin{figure}
\epsscale{1.0}
\plotone{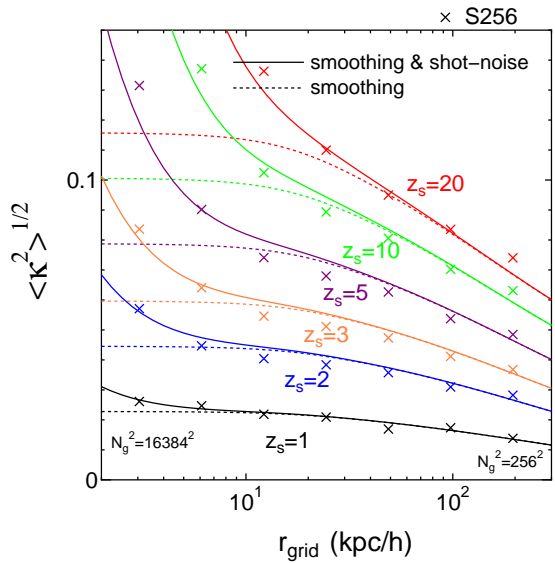}
\caption{
 Same as previous figure \ref{fig_kappa} for the model S256 (i.e., the
 lowest resolution run), but including the effects of the smoothing
 and the shot noise in the theoretical model. The solid curves are the
 theoretical model including effects of both the shot noise and the
 smoothing in Equation~(\ref{rms_kappa_v2}). The dotted curves are
 predictions with the smoothing effect only. 
}
\label{fig_kappa_v2}
\vspace*{0.5cm}
\end{figure}

To understand the effects of the shot noise and the smoothing, we
rewrite Equation~(\ref{rms_kappa}) by taking these effects into
account: 
\beqa
 \langle \kappa^2 \rangle = \frac{9}{8 \pi} H_0^4 \Omega_m^2 \int_0^{z_s}
 \frac{dz}{H(z)} \left( 1+z \right)^2 \left[ \frac{r(z) r(z,z_s)}{r(z_s)}
 \right]^2  \nonumber \\
 \times \int dk k \left[ P(k,z) + \frac{1}{n} \right]
 ~e^{-(k/k_{\rm cut})^2},  
\label{rms_kappa_v2}
\eeqa
where $1/n$ is the shot noise term ($n$ is the number density of the $N$-body 
 particles defined as $n=N_p^3/L^3$) and $e^{-(k/k_{\rm cut})^2}$ is the
 Gaussian smoothing term
 with a cutoff wave number $k_{\rm cut}= \alpha \times 2 \pi/r_{\rm grid}$
 ($\alpha$ is a constant order of unity; we adopt $\alpha=0.2$). 
 In Figure~\ref{fig_kappa_v2}, the theoretical model given by
 Equation~(\ref{rms_kappa_v2}) is compared with the lowest resolution
 ray-tracing results. We find that this model with 
 the shot noise and smoothing is in good agreement with the simulation results
 for both the small and the large $r_{\rm grid}$. For comparison,
 dotted curves show the predictions when only the smoothing term in
 Equation~(\ref{rms_kappa_v2}) is included. The differences between the
 solid and dotted curves clearly show the effect of the shot noise at
 small $r_{\rm grid}$. Furthermore, our calculation provides a way to
 infer the critical grid scale from which the shot noise dominates.
 We note that we plot the low resolution simulation results with
 $N_p^3=256^3$ to show the shot noise effect clearly. As shown in
 Figure~\ref{fig_kappa}, in the highest resolution model S1024, 
 the shot noise is negligible even at redshift $z_s=20$ and the
 smallest grid size of $3h^{-1}$kpc. In what follows, we use this
 highest resolution model to compute the PDFs for different redshifts
 and grid scales.

\subsection{Convergence, Magnification and Shear PDF}

In this section, we show our simulation results of the probability
 distribution of the convergence, magnification and shear
 up to $z_s=20$. We will show the PDFs for the highest resolution
 simulation of model S1024 with the smoothing length $3h^{-1}$kpc. 
 We have checked that the PDFs are consistent with those computed from
 lower resolution simulations, as long as the grid size is large
 enough for the shot noise not to dominate. 

 As noted in Section~\ref{sec_kappa},  the PDF in the source plane
 $dP_S/dx$ is different from that in the image plane $dP_I/dx$ by a
 factor of the magnification: 
\beq
 \frac{dP_S(x)}{dx} = \frac{1}{\mu} \frac{dP_I(x)}{dx},
\label{dp_image_source}
\eeq
where $x$ denotes the convergence $\kappa$, shear $\gamma$, and
magnification $\mu$. 

\subsubsection{Convergence PDF}

\begin{figure}
\epsscale{1.0}
\plotone{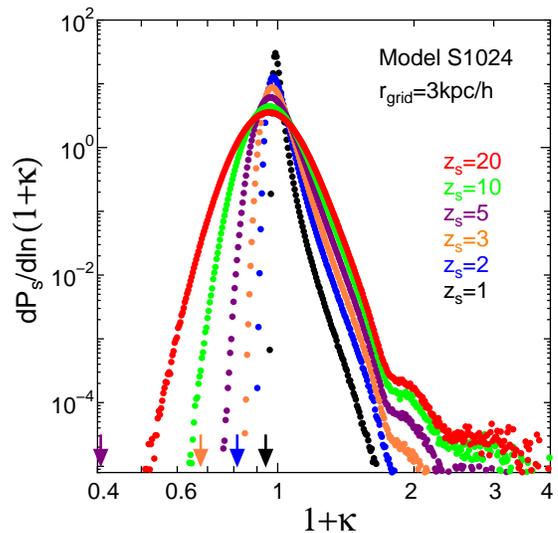}
\caption{
 Convergence PDF as a function of $1+\kappa$ for
 distant sources at $z_s=1-20$. We show the results of our highest
 resolution simulation (model S1024, the smoothing scale
 $3h^{-1}$kpc). The arrows indicate the minimum convergence for the
 empty beam.  
}
\label{fig_convPDF}
\vspace*{0.5cm}
\end{figure}

Figure~\ref{fig_convPDF} shows the convergence PDF as a function of
$1+\kappa$, not $\kappa$ itself, at redshifts $z_s=1-20$.
For higher redshifts, the distribution becomes broader and its peak moves
 to lower value.
These features are consistent with previous works (e.g., Jain et al. 2000).
With increasing the smoothing scale $r_{\rm grid}$, the distribution
 becomes narrower as expected from the variance in Figure~\ref{fig_kappa}.
As shown in the Figure, there are small dumps at $1+\kappa \simeq 2$
 because of the multiple images form for $\kappa \gtrsim 1$. 
Given the way we compute the source-plane PDFs, we count the images
more than once for a single source.
We will discuss how to treat the multiple images in
Section~\ref{sec:stronglens}.  

The convergence has a minimum value when the light ray propagates through
 the empty region (so called the empty beam, e.g. Jain et al. 2000). 
The convergence for the empty beam is given by,
\beq
  \kappa_{\rm empty} = - \frac{3}{2} H_0^2 \Omega_m \int_0^{z_s}
 \frac{dz}{H(z)} \left( 1+z \right) \frac{r(z)r(z,z_s)}{r(z_s)}.
\label{kappa_empty}
\eeq
The arrows in Figure~\ref{fig_convPDF} show the minimum convergence of
the empty beam, $1+\kappa_{\rm empty}$.
As shown in the Figure, for a low source redshift (e.g. $z_s=1$),
 the minimum convergence of the empty beam (the arrow) is consistent
 with the minimum value of the PDF. However, for higher redshifts, 
 the convergence of the empty beam becomes significantly lower than 
 the minimum values of the PDFs. This is because the matter
 distribution becomes more homogeneous for higher redshift universe
 and hence there is little chance to propagate through the empty
 region. 

\begin{figure}
\vspace*{0.5cm}
\epsscale{1.0}
\plotone{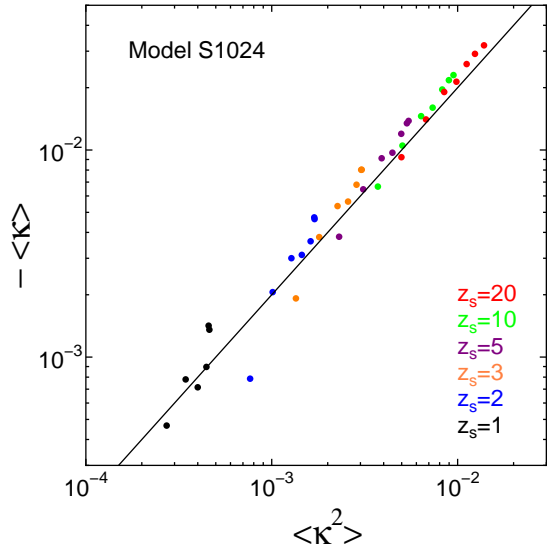}
\caption{
The negative mean of the convergence as a function of the variance of
the convergence. The dots are our simulation results for various
smoothing scales 
 $r_{\rm grid}=3-200h^{-1}$kpc and source redshifts $z_s=1-20$.
The solid line is a simple theoretical prediction, $\langle \kappa \rangle =
 - 2 \langle \kappa^2 \rangle$ (see text for details).
The mean convergence has a strong correlation with its variance, which is
explained well by the simple relation above. 
}
\label{fig_mag_mean}
\vspace*{0.5cm}
\end{figure}

In order to check our numerical simulations, we evaluate the mean of
 the magnification $\langle \mu \rangle$ and the convergence
 $\langle \kappa \rangle$.
We find that the mean magnification is $\langle \mu \rangle =1$
 within small scatters less than $1 \%$ for $z_s=1-20$.
Hence, by averaging many light rays, we recover the filled-beam
distance assuming the homogeneous mass distribution in the standard
cosmological model. On the other hand, we find that the mean convergences
$\langle \kappa \rangle$ in ray-tracing simulations are not zero but
systematically smaller than zero. Figure~\ref{fig_mag_mean} shows our
numerical results of the mean convergence $\langle \kappa \rangle$ as
a function of the variance $\langle \kappa^2 \rangle$. We show the
results for various smoothing scales $r_{\rm grid}=3-200h^{-1}$kpc and
source redshifts $z_s=1-20$. As shown in the Figure, $\langle \kappa
\rangle$ is systematically negative, which is contradictory to the
naive expectation in the weak lensing limit, $\langle \kappa \rangle =
0$. We find that the mean is well correlated with the variance, which
can be understood as follow. Assuming $\kappa$ is small, we can expand
the magnification as, 
\beqa
  \mu &=& \frac{1}{(1-\kappa)^2-\gamma^2},  \nonumber   \\
      &\simeq& 1 + 2\kappa + 3\kappa^2 + \gamma^2.
\label{mean_mu_0}
\eeqa
Then, by taking the mean of the above equation, we obtain
\beq
 1 = \langle \mu \rangle \simeq 1 + 2 \langle \kappa \rangle +
 4 \langle \kappa^2 \rangle,
\label{mean_mu_1}
\eeq
from which we have $\langle \kappa \rangle \simeq -2 \langle
\kappa^2 \rangle$.  The solid line in Figure~\ref{fig_mag_mean}
corresponds to the theoretical prediction in Equation~(\ref{mean_mu_1}).
We find that the simulation results are explained well by this simple
relation. The small deviation from the solid line can be due to the
higher order  moments of the convergence and the shear neglected in
Equation~(\ref{mean_mu_0}). Our simulation indicates that the light rays pass
through underdense region on an average.
This area magnification effect on the average value of the convergence
was also derived by a perturbative approach (Hamana 2001).
A possible explanation for
this is that, when a light ray passes near a massive object, its paths
is defected such that the closest distance to the massive object
becomes larger and hence the ray passes through the less dense region. 
The negative mean of the convergence, for instance, can have an
impact on the statistics of time delays (Oguri 2007).

We note that the above results depend on which plane (image or source plane)
 we use to calculate the mean, because 
 the higher magnification events are more
 weighted in the image plane.
We find that the mean convergence in the image plane is zero
 within small scatters less than $10^{-3}$.
This is because the mean in the image plane is different by a factor of
 $\mu$ from Equation (\ref{dp_image_source}),
 and hence the mean convergence in the image plane is $\langle \kappa
 \rangle_{\rm image} = \langle \mu \kappa \rangle \simeq
 \langle (1+2\kappa) \kappa \rangle = 0$ 
 from Equation (\ref{mean_mu_1}).
Similarly, the mean magnification in the image plane is found to be
 systematically positive, $\langle \mu \rangle_{\rm image} =
 \langle \mu^2 \rangle > 0 $.
In addition, our results are derived under an ideal situation in
which light sources are distributed uniformly in a source plane and all
the sources are supposed to be observed. In general, only sources with
luminosity above a certain threshold are observed. In that case, the
mean magnification can be non-zero with either sign because of the so-called
magnification bias (e.g. Schneider et al 1992 and see section 5.2).
Hence, we have to use the appropriate statistics depending on 
 situations.

\begin{figure}
\epsscale{1.1}
\plotone{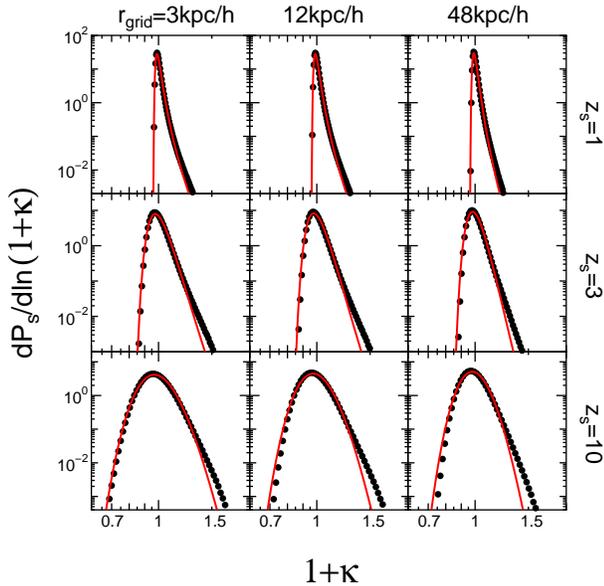}
\caption{
The convergence PDF for various smoothing scales $r_{\rm grid}$ and
 redshifts $z_s$. Black dots are our simulation results, while red solid
 curves are the modified log-normal fitting model in Das \& Ostriker (2006).
}
\label{fig_convPDF_Das}
\vspace*{0.5cm}
\end{figure}

Finally, we compare our numerical results of the convergence PDF with
previous works. Taruya et al. (2002) conducted the ray-tracing
simulations to investigate the statistical properties of weak-lensing
field. They showed that the convergence PDF is well described by
the lognormal distribution, which reflects the fact that the one-point
distribution function of matter density field is well described by the
lognormal model (Kayo et al. 2001).
Later, Das \& Ostriker (2006) calculated the probability distribution
of the projected surface mass density using cosmological $N$-body
 simulations to show that the PDF is well fitted by the modified
 lognormal distribution:
\beqa
 \frac{dP_s(\kappa)}{d\kappa} = N_\kappa 
 \exp \left[ -\frac{1}{2 \omega^2_\kappa}  
 \left\{ \ln \left( 1+ \frac{\kappa}{\left| \kappa_{\rm empty} \right|}
 \right) + \frac{\omega^2_\kappa}{2} \right\}^2 \right. \nonumber \\
 \left. \times \left\{ 1+\frac{A_\kappa}{1+ \kappa / \left| \kappa_{\rm empty}
 \right|} \right\} \right] \frac{1}{\kappa+\left| \kappa_{\rm empty}
 \right|},~~~~~
\label{lognormal_Das}
\eeqa
where $N_\kappa$ is the normalization. By setting $A_\kappa=0$ and
$\omega_\kappa=\langle \kappa^2 \rangle$,
Equation~(\ref{lognormal_Das}) reduces to the lognormal model of
Taruya et al. (2002). The two parameters $A_\kappa$ and $\omega_\kappa$ are
determined by using the following conditions, 
\beqa
 \int_{\kappa_{\rm empty}}^\infty d\kappa \frac{dP_s}{d\kappa} \kappa
 = \langle \kappa \rangle = -2 \langle \kappa^2 \rangle, \nonumber \\ 
  \int_{\kappa_{\rm empty}}^\infty d\kappa \frac{dP_s}{d\kappa}
 \kappa^2 = \langle \kappa^2 \rangle.
\label{Das_params}
\eeqa
We use our simulation results of $\langle \kappa^2 \rangle$ in
the above equations.
Note that the first condition of Equation~(\ref{Das_params}) differs
from the original condition used in Das \& Ostriker (2006), who
adopted $\langle \kappa \rangle=0$. We modify this condition to the
form above based on our finding shown in Figure~\ref{fig_mag_mean} and
Equation~(\ref{mean_mu_1}). 

Figure~\ref{fig_convPDF_Das} is the comparison of the convergence PDF
between our simulation results and the model of Das \& Ostriker (2006). 
Black dots are our simulation results, while red curves are their
fitting function. The nine panels are for various smoothing scales of
 $r_{\rm grid}=3h^{-1}$kpc (left), $12h^{-1}$kpc (middle) and
 $48h^{-1}$kpc (right) and for
 various source redshifts of $z_s=1$ (top), $3$ (middle) and
 $10$ (bottom).
As seen in the Figure, the fitting model agrees well with our results
in all the panels, especially near the peak of the distribution.
However, for high convergence tail, the model slightly underestimates
the PDFs

\subsection{Magnification PDF}

\begin{figure}
\vspace*{0.5cm}
\epsscale{1.}
\plotone{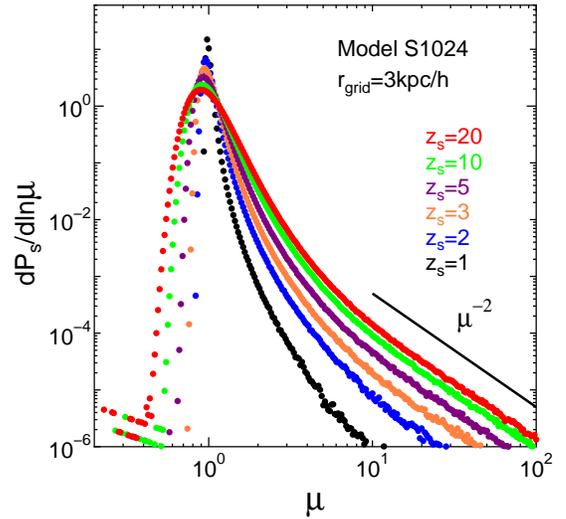}
\caption{
Magnification PDFs of the distant sources at $z_s=1-20$.
We show the results of our highest resolution simulation
 (model S1024, the smoothing scale $3h^{-1}$kpc).
}
\label{fig_magPDF}
\vspace*{0.5cm}
\end{figure}

Figure~\ref{fig_magPDF} shows the magnification PDF of the distant
sources at $z_s=1-20$. 
We use the highest resolution model S1024 (see Table~\ref{table1}) with
 the smoothing scale $3h^{-1}$kpc. As clearly seen in the Figure, for
 more distant sources, the peak of PDF moves to the fainter magnification
 (less than $1$) and its distribution becomes broader as in
 Figure~\ref{fig_convPDF}.
These features
 are consistent with the previous works (e.g., Wambsganss et al. 1998;
 Wang et al. 2002; Hilbert et al. 2007). For example, at the highest
 redshift $z_s=20$, $10 \% (1 \%)$ sources are magnified by over $30\%
 (100\%)$. For high magnification ($\mu>10$), the simulation results
 are asymptotically proportional to $\mu^{-2}$ which is consistent
 with the analytical expectation (e.g., Schneider et al. 1992). We note
 that there are small increases of PDFs at very low magnification
 ($\mu < 1$). These correspond to fainter images of strongly lensed
 multiple images. Again, we discuss strong lensing events in more
 detail in Section~\ref{sec:stronglens}. 

\begin{figure}
\epsscale{0.9}
\plotone{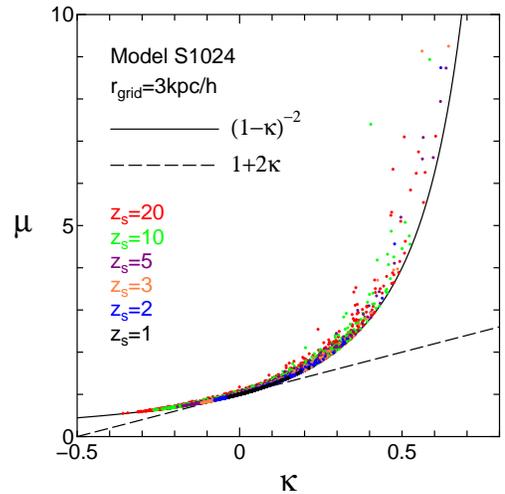}
\caption{
The two dimensional distribution of the convergence and the magnification.
The solid curve is $\mu=(1-\kappa)^{-2}$ and the dashed line is
 $\mu=1+2 \kappa$ (weak lens approximation). 
}
\label{fig_2d_magnf_conv}
\vspace*{0.5cm}
\end{figure}

\begin{figure}
\vspace*{0.5cm}
\epsscale{1.1}
\plotone{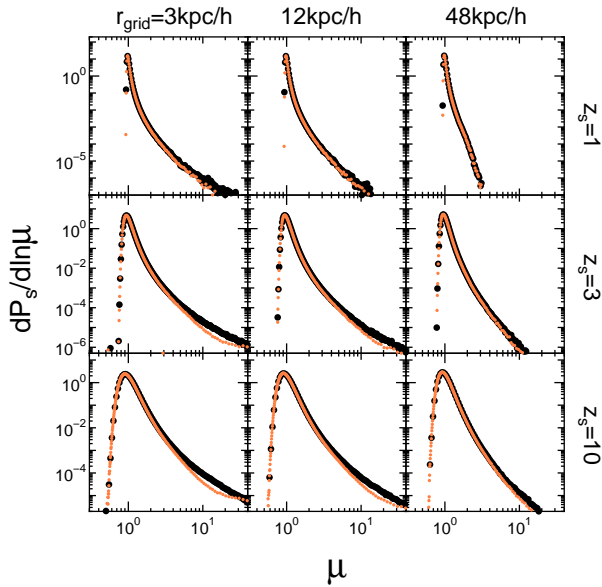}
\caption{
The magnification PDFs for various smoothing scales $r_{\rm grid}$ and
 redshifts $z_s$. Black dots are our simulation results. Orange dots
 denote the magnification PDFs converted from the convergence PDFs in
 our simulations via the relation $\mu=(1-\kappa)^{-2}$. 
}
\label{fig_magPDF_Oguri}
\vspace*{0.5cm}
\end{figure}

\begin{figure}
\vspace*{0.5cm}
\epsscale{1.1}
\plotone{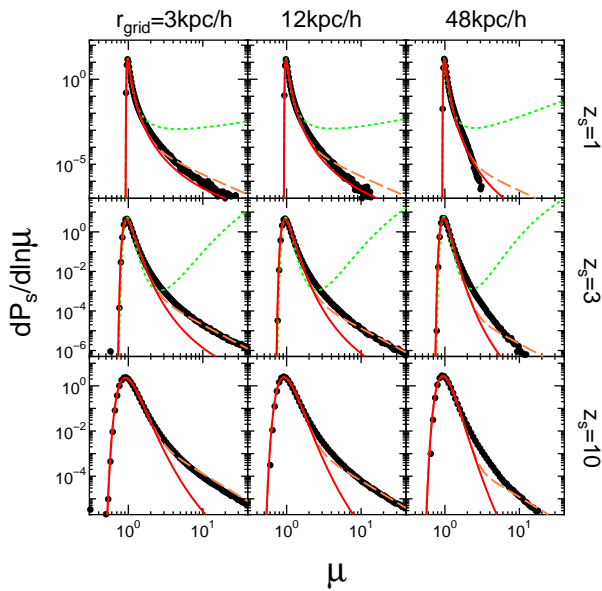}
\caption{
The magnification PDF for various smoothing scales $r_{\rm grid}$ and
 redshifts $z_s$. Black dots shows our simulation results, green dotted
 curves are the fitting model in Wang et al. (2002), red solid curves are
 the fitting function in Das \& Ostriker (2006) converted via the
 relation $\mu=(1-\kappa)^{-2}$, orange dashed curves are same as the
 red solid curves but including the high magnification tail.  
}
\label{fig_magPDF_Das}
\vspace*{0.5cm}
\end{figure}

In previous works, the convergence and magnification PDFs have been
presented separately. Here we investigate the relation of these two
PDFs. First we check the correlation of the convergence $\kappa$ and
magnification $\mu$, which is shown in Figure~\ref{fig_2d_magnf_conv}.
The vertical axis is $\mu$, while the horizontal axis is $\kappa$, for
redshifts $z_s=1-20$. The Figure indicates that convergence and magnification
are highly correlated with each other, which is consistent with Barber
et al. (2000) and Hilbert et al. (2011). We find that the correlation
is well explained by $\mu=(1-\kappa)^{-2}$ (solid curve), which comes
from the definition of the magnification with the shear term neglected.
On the other hand, the relation $\mu=1+2 \kappa$ (dashed line), which
holds in the weak lensing approximation, cannot explain the
correlation in our simulations very well. 

Given the tight correlation, we translate the convergence PDF to the
magnification PDF using the relation
$\mu=(1-\kappa)^{-2}$. Specifically, we compute the magnification PDF
as  
\beq
 \frac{dP_s}{d\mu} = \frac{(1-\kappa)^3}{2} \frac{dP_s}{d\kappa}.
\label{convPDF_magPDF}
\eeq
The result is shown in Figure~\ref{fig_magPDF_Oguri}. The black dots are
the original magnification PDFs, while the orange dots are those
converted from the convergence PDFs in our simulations using the
relation  $\mu=(1-\kappa)^{-2}$. Note that we use the simulated
convergence PDFs only for $\kappa<1$, as the conversion equation
clearly breaks down at $\kappa=1$. We find that these two PDFs agree
surprisingly well, up to very high magnification tails of $\mu\sim
10$. Hence, once either the convergence or magnification PDF is
available, we can easily obtain the other PDF by the transformation
given above. 

We now compare our numerical results of magnification PDF with
previous works. Wang et al. (2002)  proposed the fitting function of
the magnification PDF using simulation results of Wambsganss et
al. (1997), Barber et al. (2000) and Munshi \& Jain (2000). Their
formula depends only on the variance of the convergence, and
independent of cosmological models and redshifts. Their fitting
function is a stretched Gaussian distribution. In
Figure~\ref{fig_magPDF_Das}, green dotted curves show their model, while
black dots are our simulation results. We find that their model agree
well only near the peak of the distribution.  In fact, as they
noticed, their model can be used only for near the peak $\mu \simeq 1$.
They used the approximation of $\mu=1+2\kappa$ and their model depends
on the variable $\mu_{\rm min}=1+2 \kappa_{\rm empty}$. 
However, for high redshift $z_s \gtrsim 4$, this quantity $\mu_{\rm min}$
becomes negative, leading to the break down of the formula (hence we
do not plot the green dotted curves for $z_s=10$ in
Figure~\ref{fig_magPDF_Das}). 

Since the modified lognormal model in Das \& Ostriker (2006) well
reproduces the convergence PDF, we convert this model to the
magnification PDF using Equation~(\ref{convPDF_magPDF}) to see if the
model reproduces the simulated magnification PDFs.
We find that this model (red curves in Figure~\ref{fig_magPDF_Das})
agrees reasonably well with our simulation results, although the model
cannot reproduce the large magnification tail, presumably reflecting
the slight underestimate of the convergence PDFs at the high
convergence tail. Since we know that the magnification PDF behaves as
$dP_s/d\mu \propto \mu^{-3}$ at the tail, we propose a simple
phenomenological model which takes account of the tail behavior,  
\beqa
  \frac{dP_s}{d\mu} = \frac{(1-\kappa)^3}{2} \frac{dP_s}{d\kappa}
 + \Theta(\mu-1) \exp \left[ -\frac{1}{4 \left( \mu-1 \right)^4} \right]
  \nonumber \\
 \times \frac{(1-\kappa_0)^3}{2} \left. \frac{dP_s}{d\kappa}
 \right|_{\mu=\mu_0}  \left( \frac{\mu}{\mu_0} \right)^{-3}.
\label{Das_Oguri_eq}
\eeqa
where $\Theta$ is the step function and $\mu_0=(1-\kappa_0)^{-2}$. We
set $\mu_0=3$ as a reasonable choice to fit our simulation results for
a wide range of parameters. In right hand side of
Equation~(\ref{Das_Oguri_eq}), the convergence PDF $dP_s/d\kappa$ is
evaluated by using the modified lognormal model of Das \& Ostriker (2006). 
This model, shown by the orange dashed curves in Figure~\ref{fig_magPDF_Das},
reproduces the simulation results well both near peaks and at high
magnification tails. We note that a more robust and accurate model of
the magnification PDF can be obtained by computing the tail
distribution separately using the halo model to combine it with the peak
distribution computed from the modified long-normal convergence PDF.

\subsection{Shear PDF}


\begin{figure}
\epsscale{1.0}
\plotone{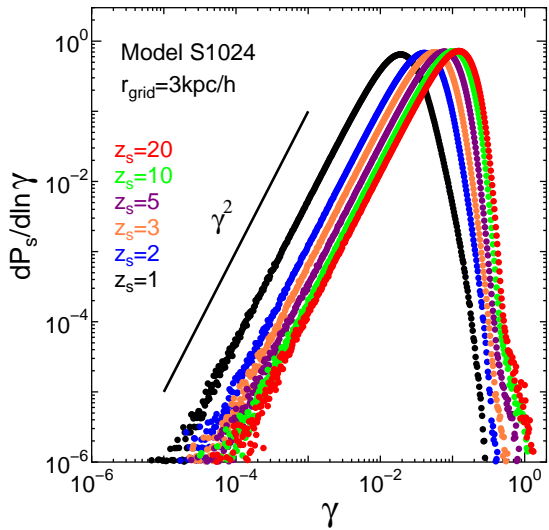}
\caption{
The shear PDF of distant sources at $z_s=1-20$. 
The results are obtained from our highest resolution simulation
(model S1024, the smoothing scale $3h^{-1}$kpc).
}
\label{fig_shearPDF}
\vspace*{0.5cm}
\end{figure}

\begin{figure}
\epsscale{1.1}
\plotone{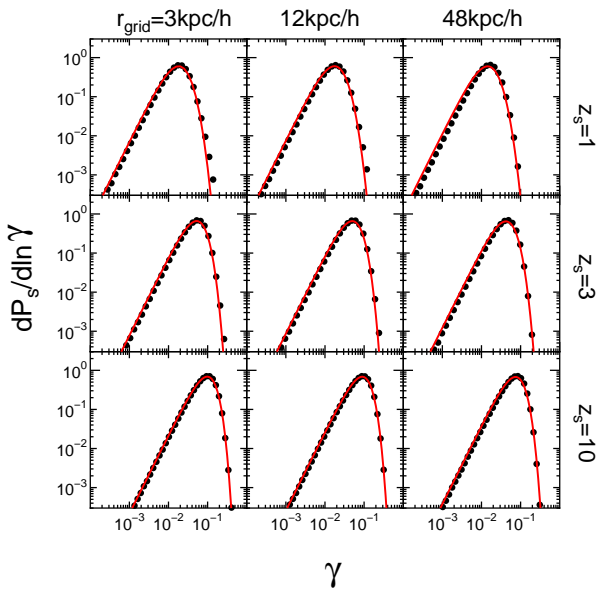}
\caption{
The shear PDF for various smoothing scales $r_{\rm grid}$ and redshifts
 $z_s$. Black dots are our simulation results, while red curves are
 our fitting formula in Equation~(\ref{shearPDF_oguri}). 
 Our fitting formula agrees with the data very well.
}
\label{fig_shear_oguri_PDF}
\vspace*{0.5cm}
\end{figure}


Figure~\ref{fig_shearPDF} shows the shear PDF as a function of the shear
 amplitude $\gamma=(\gamma_1^2+\gamma_2^2)^{1/2}$. 
With increasing the redshift, the peak of the PDF becomes larger
while its shape is not changed very much. The shear PDF is
proportional to $\gamma^2$ for the small shear amplitude ($\gamma \ll
1$).  This behavior has been explained by an analytical calculation for
light rays propagating through randomly distributed point-mass lenses
 (e.g., Schneider et al. 1992, Sec. 11.2). The tidal field of the lenses
 generates the shear, while the convergence is exactly zero because
 the light propagates the empty space between the point-mass lens
 particles. The analytical PDF is $dP_s/d\ln\gamma \propto \gamma^2$ for
 $\gamma \ll \kappa_*$ and $\propto \gamma^{-1}$ for
 $\gamma \gg \kappa_*$ where $\kappa_*$ is the convergence for
 the surface density of the uniformly distributed point-mass lenses.
 Our simulation result is consistent with the simple theoretical model
 only for small $\gamma$. 

Here we present a fitting formula for the shear PDF.
We empirically derives the following fitting formula for the PDF of
 $\gamma$,
\beq
 \frac{dP_s(\gamma)}{d\gamma} = N_\gamma 
  \frac{\gamma}{\omega^2_\gamma}  \exp \left[
  - \frac{\left\{ \ln \left( 1 + A_\gamma \gamma \right) \right\}^2}{A^2_\gamma \omega^2_\gamma}
 \right],
\label{shearPDF_oguri}
\eeq
where $N_\gamma$ is a normalization.
In order to determine the parameters $A_\gamma$ and $\omega_\gamma$, we use 
 $18$ shear PDFs for three different smoothing scales $r_{\rm grid}=3,
 12,48h^{-1}$Mpc with six redshifts $z_s=1,2,3,5,10$ and $20$.
 We find the best-fitting values of $A_\gamma$ and $\omega_\gamma$ are well
 correlated with the shear variance 
$\sigma_\gamma (=\langle \gamma^2 \rangle^{1/2})$ as
\beq
  A_\gamma = \frac{0.010}{\sigma_\gamma \left( 0.00075 + \sigma_\gamma^2
 \right)^{0.7}} ,
 ~~\omega_\gamma = 1.66 ~\sigma_\gamma^{1.32}.
\label{shearPDF_oguri_param}
\eeq
Figure~\ref{fig_shear_oguri_PDF} shows the above fitting formula (red curves)
 and the simulation results (black dots) for various smoothing scales
 $r_{\rm grid}$ and redshifts $z_s$.
As shown in the Figure, our fitting formula agrees with the simulation
results quite well for a wide range of redshifts and the smoothing scales.

Finally, we investigate the correlations between shear and convergence.
A correlation coefficient $r_{xy}$ between the quantities $x$ and $y$
 is defined as,
\beq
 r_{xy} = \frac{\langle (x-\langle x \rangle) 
 (y-\langle y \rangle) \rangle}{\langle (x-\langle x \rangle)^2
 \rangle^{1/2} \langle (y-\langle y \rangle)^2 \rangle^{1/2}},
\label{corr_coef}
\eeq  
where $x,y$ represents $\kappa$, $\gamma_{1,2}$ and $\gamma$.

\begin{figure}
\epsscale{1.0}
\plotone{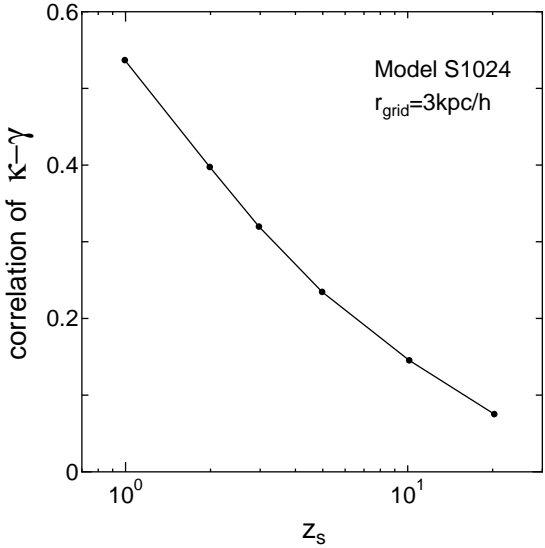}
\caption{
Correlation between $\gamma$ and $\kappa$ as a function of
the source redshift. The vertical axis is the correlation coefficient
defined in Equation~(\ref{corr_coef}).
}
\label{fig_corr_kappa_gamma}
\vspace*{0.5cm}
\end{figure}

We find a positive correlation between $\kappa$ and $\gamma$, while we
find no correlations among $\gamma_1$, $\gamma_2$, and $\kappa$. 
Figure~\ref{fig_corr_kappa_gamma} shows the correlation between
 $\kappa$ and $\gamma$. The vertical axis is the correlation
 coefficient $r_{\kappa,\gamma}$ defined in Equation~(\ref{corr_coef}),
and the horizontal axis is the source redshift. 
The Figure shows the strong correlation of $0.1-0.5$ with a clear
tendency that the correlation is stronger for lower redshift sources.

The positive correlation between $\kappa$ and $\gamma$ is expected at
least when a single lensing event near a halo is dominant. On the
other hand, for multiple lensing by different halos, the total
convergence is simply given by the sum of each event, whereas the
shear can be cancelled out depending on the relative direction of the
shear in each event. This implies that the correlation becomes weaker
when there are more intervening halos that can potentially contribute
to lensing. This simple consideration appears to be consistent with
the simulated results above, as higher redshift sources have more
intervening halos that can significantly affect the light
propagation.

\section{Strong lensing probability}
\label{sec:stronglens}

\begin{figure}
\epsscale{1.0}
\plotone{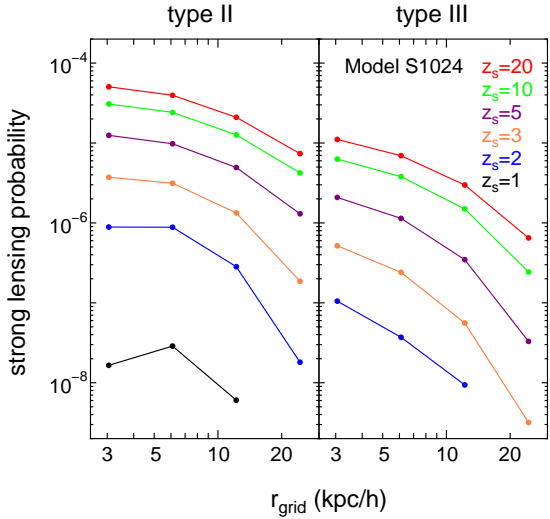}
\caption{
Strong lensing probability for type II (left panel) and type III (right
 panel).
}
\label{fig_stlp}
\vspace*{0.5cm}
\end{figure}

\begin{figure*}
\vspace*{0.5cm}
\epsscale{1.1}
\plotone{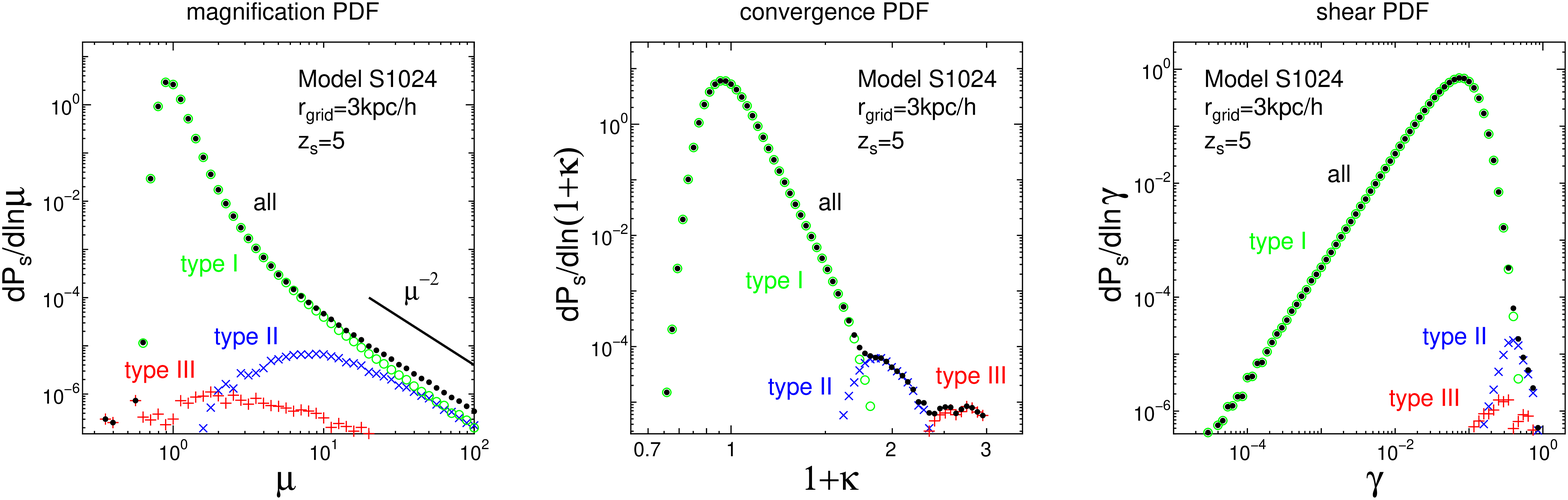}
\caption{
Magnification PDF (left panel), convergence PDF (middle panel) and shear
 PDF (right panel) for $z_s=5$ for different types of lensing mapping:
green circles are type I, blue crosses are type II, red pluses
 are type III, and black dots are the sum of them.
}
\label{fig_magPDF_lens}
\vspace*{0.5cm}
\end{figure*}

In this section, we calculate the strong lensing probability and
 investigate the effects of the multiple images on the PDFs.

The light-ray path from the source to the observer is the stationary
point of the time delay $\tau$ (or the Fermat potential), i.e.,
 $\partial \tau / \partial \theta_i=0$, due to the Fermat's principle.
The lensing mapping from the image position to the source position is 
 characterized using the Jacobian matrix which is the Hessian of $\tau$, 
 $A_{ij}=\partial^2 \tau / \partial \theta_i \partial \theta_j$.
Using the convergence $\kappa$ and the shear $\gamma_{1,2}$, the
 Jacobian matrix is expressed as (e.g., Schneider et al. 1992)
\beq
  A = \left(
     \begin{array}{@{\,}cc@{\,}}
        1-\kappa-\gamma_1 & -\gamma_2 \\
        -\gamma_2 & 1-\kappa+\gamma_1  
     \end{array}
 \right).    \nonumber
\eeq
The mapping is categorized into three types: the minimum, the maximum,
 and the saddle point of $\tau$.
The minimum point is called type I, the saddle point is type II, and
 the maximum point is type III (e.g., Schneider et al. 1992): 
\beqa
 \rm{type ~I~} &&: ~{\rm det}A > 0 ~\&~ {\rm tr}A > 0,  \nonumber  \\
               &&~~~~{\rm i.e.},~ | 1-\kappa | > \gamma ~\&~  
                   \kappa < 1 \nonumber  \\
 \rm{type ~II~} &&: ~{\rm det}A < 0, ~~{\rm i.e.},~
                     | 1-\kappa | < \gamma    \\
 \rm{type ~III~} &&: ~{\rm det}A > 0 ~\&~ {\rm tr}A < 0,  \nonumber  \\  
                 &&~~~~{\rm i.e.},~ | 1-\kappa | > \gamma ~\&~  
                   \kappa > 1 \nonumber 
\eeqa
Type II and III correspond to multiple images produced by strong lensing.
The strong lensing probability is defined as the ratio of the number of
 light rays in theses two types,
\beq
  P_{\rm II,III}=\frac{\sum_i (1/\mu_i)_{\rm type II, III}} 
     {\sum_i (1/\mu_i)_{\rm all types}}.
\eeq
The factor $1/\mu_i$ is due to the probability in the source plane.
Figure~\ref{fig_stlp} shows the strong lensing probability of type II (left
 panel) and type III (right panel) as a function of $r_{\rm grid}$.
As shown in the Figure, as the smoothing scale decreases, the
probability increases. This is because the smoothing effect smears out
the central cusp of the dark halo.  The probability for type II is
higher than that for type III. 

Figure~\ref{fig_magPDF_lens} shows the contribution of each type
 to the PDFs at $z_s=5$.
The left panel is the magnification PDF, the middle panel is
 the convergence PDF, and the right panel is the shear PDF.
The green circles are type I, the blue crosses are type II, the red
 pluses are type III, and the black dots are the sum of them.
As clearly seen in the Figure, almost all light rays are type I.
With increasing the source redshift, the fractions of type II and III
increase. In the magnification PDF, the small increase seen at $\mu
\lesssim 0.6$ is due to type III.  In the high magnification limit,
type I and II have same probability. In the convergence PDF, the small
knot at $1+\kappa \simeq 2$ is type II, and the second knot at
$1+\kappa \simeq 3$ is type III.
In the shear PDF, type II and III appear in the high shear limit. 
 The resolution of our simulation is
high enough to see these strong lensing features clearly in the PDFs.

In order to understand these three types more clearly, we consider the
strong lensing by a NFW halo (Navarro, Frenk \& White 1997). For the NFW
profile, the caustic is the circle on the source plane. If the source
position is outside the caustic,  only one image of type I is formed.
When the source position crosses the caustic inward, two new images
are formed. One is formed the opposite side of the lens center, which
is type II, and the other one is a faint image formed close to the
center, which is type III. Hence, as shown in
Figure~\ref{fig_magPDF_lens}, the convergence of the type III is very
high $\kappa > 1$ while the magnification is small. 
When the source is close to the lens center, the two bright images of
 type I and II are strongly magnified with the similar magnification,
 which can explain the high magnification tail in the left panel of
 Figure~\ref{fig_magPDF_lens}.

\section{Magnification Effects on Luminosity Functions of Distant Sources}
\label{sec:mag}
\subsection{Effects of Source Size on Magnification PDF}

\begin{figure}
\epsscale{1.0}
\plotone{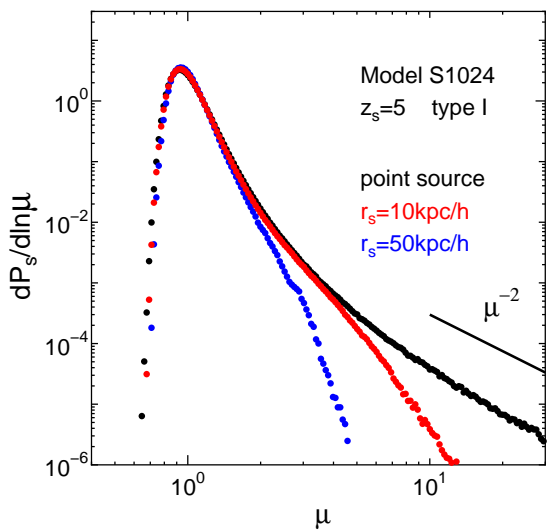}
\caption{
The magnification PDF for the extended source at $z_s=5$.
The red (blue) symbols are the results for the source radius of
 10 (50) $h^{-1}$kpc,
while the black symbols are for the point sources.
Here, we do not include the multiple images. 
}
\label{fig_magnfPDF_smooth}
\vspace*{0.5cm}
\end{figure}

So far we have discussed lensing effects for point sources.
In this section, we investigate the effects of the finite source size
on the magnification PDF. The finite size smooths the magnification
profile on the source plane, and smears out some high magnification
events ($\mu \gg 1$). 
The smooth magnification $\bar{\mu}$ of the extended source 
 with surface brightness profile $I(\eta)$ is generally given by,
\beq
\bar{\mu} = \int d^2 \eta ~\mu(\bfeta) I(\bfeta),
\label{mean_mu}
\eeq
where $\bfeta=(\eta_1,\eta_2)$ is the two-dimensional vector in the
 source plane. For simplicity, we assume that the surface brightness
 is a circular top-hat model with the radius $R_S$, $I(\bfeta)=1/(\pi
 R_S^2)$  for $\eta \leq R_S$ and $I(\bfeta)=0$ for $\eta > R_S$. 
 Note that the distances $\bfeta, R_S$ are the comoving scale. We adopt
 $R_S=10h^{-1}$kpc as a typical size of galaxies. 

In order to resolve the sub-galactic scales ($<10h^{-1}$kpc) on the
 source plane, we prepare new ray-tracing simulations with higher angular
 resolution of $0.1$ arcsec by narrowing the field-of-view with
 the fixed number of light-rays.
The field-of-view is set to be $4 \times 4$ ${\rm{arcmin}}^2$, the number
 of light-rays is $2048^2$, and $100$ realizations are prepared.
The resulting angular resolution is $4 {\rm arcmin}/2048$ $=0.1$~arcsec. 
There are $\sim 15$ light-rays on the circle with the radius of
 $10h^{-1}$~kpc even at the highest redshift $z_s=20$, which is
 sufficient to calculate the average of the magnification given by
 Equation~(\ref{mean_mu}).   

In this section, we do not include the multiple images formed by strong
 lensing in the magnification PDF, i.e., we use the only rays with
 type I. If there is a source including type II or III, we do not
 include the sample in our analysis. However, since almost all rays
 are type I, our results are not significantly changed.

Figure~\ref{fig_magnfPDF_smooth} shows the magnification PDF for various
 source radius at $z_s=5$. The three symbols for the point source
 (black), $R_S = 10h^{-1}$ kpc (red), and $R_S = 50h^{-1}$ kpc (blue).
We find that the high magnification tail significantly decreases with
increasing the source size, which demonstrate the importance of the source
size effect for high magnification events.

\subsection{Magnification Effect on Luminosity Functions}

The lensing magnification changes the observed flux or luminosity by a
 factor of $\mu$. Specifically, the observed  (lensed) luminosity
 $L_{\rm obs}$  and the unlensed luminosity $L$ are related as $L_{\rm
   obs}=\mu L$. Since the number density of the source is conserved
 with and without lensing magnification, we have
 $\Phi^L_{\rm obs}(L_{\rm obs}) dL_{\rm obs}=\Phi(L) dL$, where
 $\Phi^L_{\rm obs} ~( \Phi )$ is the luminosity function (hereafter LF)
 of sources with (without) the lensing. Thus the lensed LF is written as
\beq
  \Phi^L_{\rm obs}(L_{\rm obs}) = \int d\mu \frac{1}{\mu}
 \frac{dP_s(\mu)}{d\mu} \Phi(L_{\rm obs}/\mu).
\label{lensSchLF}
\eeq 
We consider LFs of the Schechter model for galaxies
and  the double power law model for quasars to demonstrate the
impact of lensing magnifications on observed LF. 

\begin{figure*}
\epsscale{1.0}
\plottwo{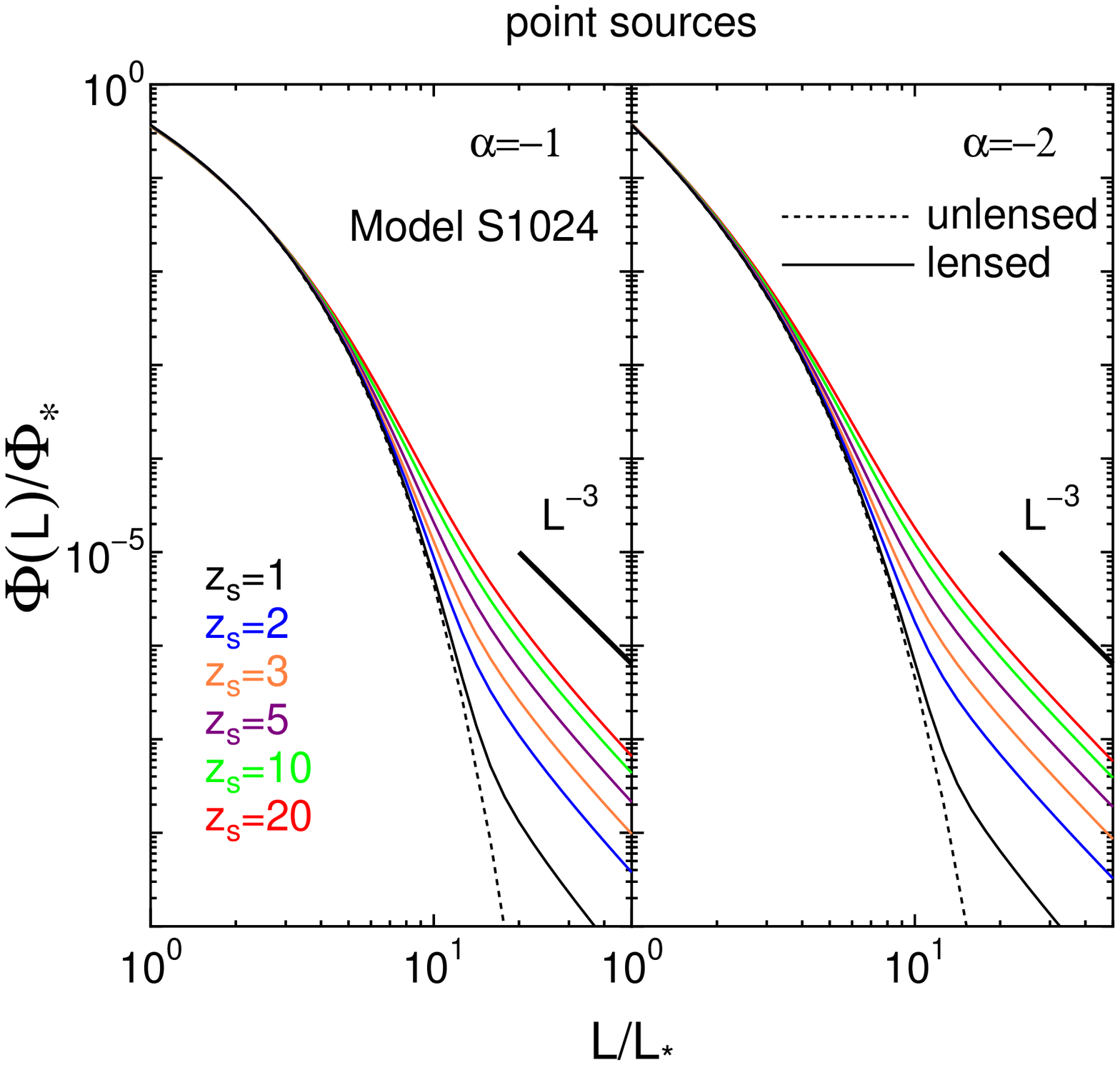}{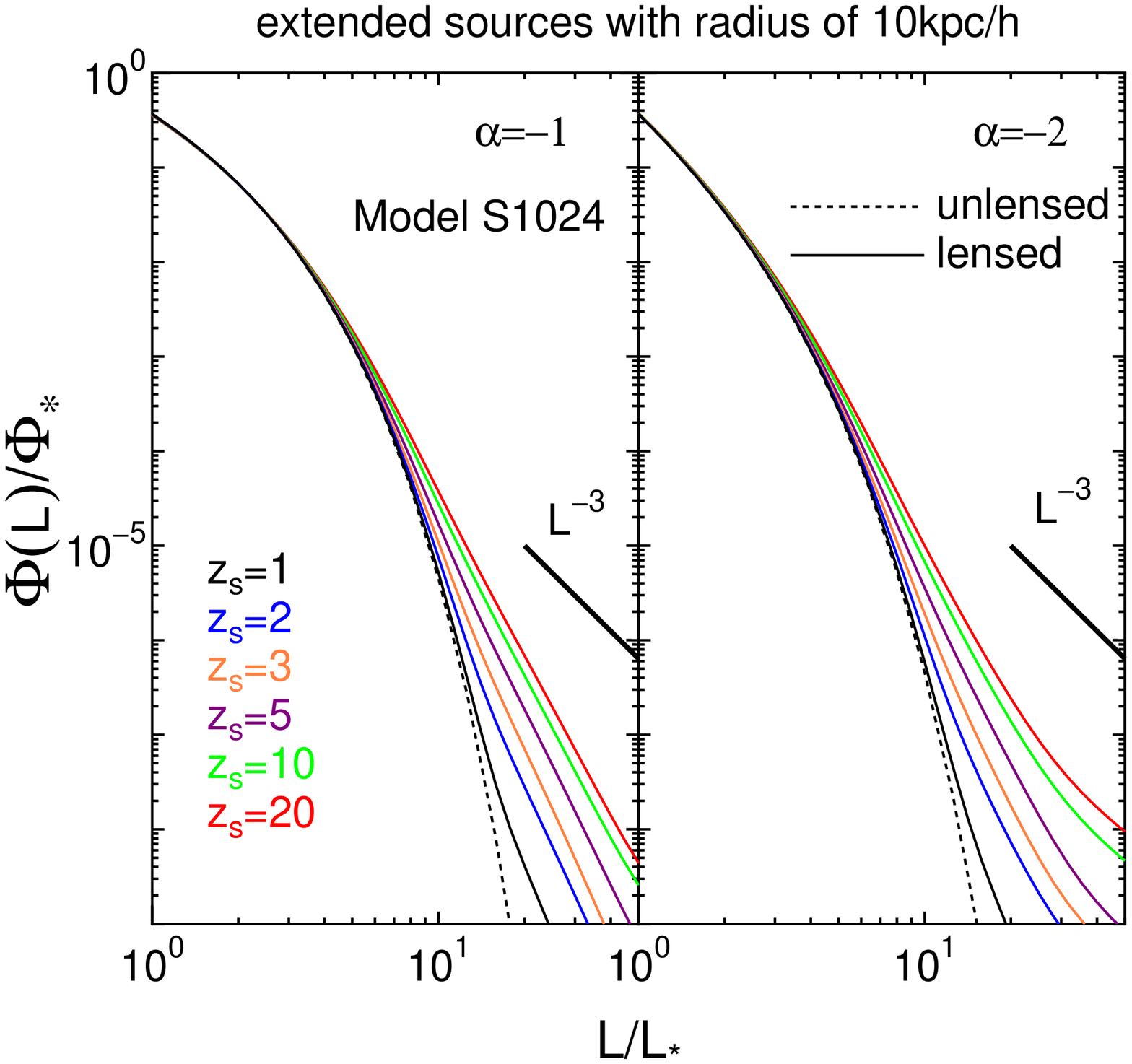}
\caption{
The lensed and unlensed Schechter luminosity function for redshifts
 $z_s=1-20$ are denoted as solid and dashed curves, respectively.
The left panel is for the point sources, while the right panel is for 
 the extended sources with the radius of $10h^{-1}$kpc (in comoving scale). 
The left (right) side in each panel is for the faint end slope $\alpha=-1$ $(-2)$.
}
\label{fig_lensSchLF}
\vspace*{0.5cm}
\end{figure*}

\begin{figure}
\epsscale{1.0}
\plotone{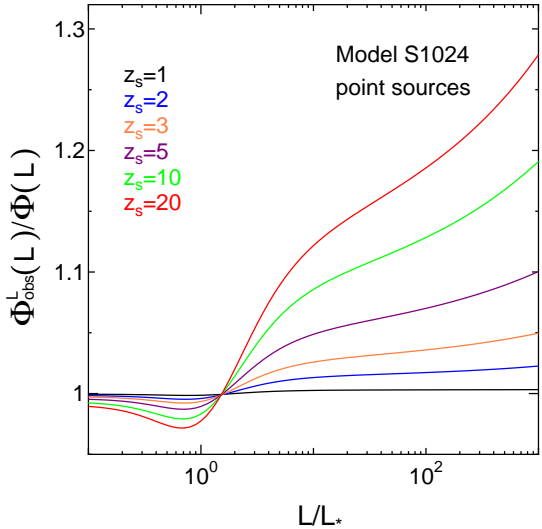}
\caption{
The lensed quasar luminosity function.
The vertical axis is the lensed model divided by the unlensed one.
}
\label{fig_lensQSO}
\vspace*{0.5cm}
\end{figure}

The Schechter LF is defined as (Schechter 1976),
\beq
\Phi(L)=\Phi_* \left( L/L_* \right)^\alpha \exp \left( -L/L_* \right),
\label{Schechter_LF}
\eeq
where $\alpha$ is the faint end slope, $L_*$ is the characteristic
luminosity, and $\Phi_*$ is the normalization. Figure~\ref{fig_lensSchLF}
shows the lensed Schechter LF for redshifts $z_s=1-20$. The solid
curves are the lensed model, while the dotted curve is the unlensed model.
Left panel is for point sources, while right panel is for extended sources
 with the radius of $10h^{-1}$~kpc. The left (right) side in each
 panel is for the faint end slope $\alpha=-1 ~(-2)$.  We can clearly
 see the lensing effect on the Schechter LF at the bright end, the
 exponential tail is modified to the power law. This is because
 there are many faint sources, some of which are magnified.
 This feature is consistent with previous works (e.g., Lima et
 al. 2010b; Wyithe et al. 2011).  As shown in the Figure, the lensed
 LF is proportional to $L^{-3}$ for the bright end $L \gg L_*$ for the
 point sources. We can understand this asymptotic behavior
 analytically by inserting $dP_s(\mu)/d\mu \propto \mu^{-3}$ (valid for
 $\mu \gg 1$)  to the Equation~(\ref{lensSchLF}) with
 Equation~(\ref{Schechter_LF}).
Then we have $\Phi_{\rm obs} \propto L_{\rm obs}^{-3}$, 
 which is independent of $\alpha$ as long as $\alpha > -3$ is satisfied.
For the finite source size, the lensing effects become less significant.   
The lensing effect appears for bright galaxies $L \gtrsim 10 L_*$,
suggesting that wide area surveys of high redshift sources are
necessary to see the lensing feature in the LF.

The double power law model for the quasar LF is
\beq
 \Phi(L) = \frac{\Phi_*}{\left( L/L_* \right)^\alpha
               +\left( L/L_* \right)^\beta},
\eeq
where $\alpha$ and $\beta$ are the slope for bright and faint sources.
We set $\alpha=3.3$ and $\beta=1.4$ to be consistent with the quasar LF
 from the Sloan Digital Sky Survey (Croom et al. 2009).
Figure~\ref{fig_lensQSO} shows the lensed LF, the ratio of lensed to
unlensed LFs. We find that the number of bright sources increase by
ten percents or so, which is not so pronounced compared with the case
of the Schechter LF. This is simply because the magnification effect is
more significant for steeper LFs. 

\section{Summary and Discussion}
\label{sec:summary}

We have presented high-resolution ray-tracing simulations to derive
accurate PDFs of the lensing convergence, shear and
magnification for distant objects. The resolution of our
simulations, for instance the softening length of the $N$-body
simulations of $2 h^{-1}$kpc (comoving scale) and the grid size of the
two-dimensional gravitational potential on the lens planes of 
$3h^{-1}$kpc (comoving scale), is high enough to enable direct 
predictions of lensing effects on distant galaxies. In addition, we
study the PDFs up to very high source redshifts of $z_s=20$.
In our $N$-body simulations, we used the different number of particles
of $256^3, 512^3$ and $1024^3$ with the fixed box size of $50h^{-1}$Mpc,
in order to address the effect of the numerical resolution carefully.
We have found that the sufficient number density of $N$-body particles
is necessary to reduce the shot noise effect which artificially
broadens the PDFs especially for high-redshift sources. Both
numerically and analytically we confirmed that our highest resolution
run with the smallest grid size of $3h^{-1}$kpc is not affected by the
shot noise effect.  
We also examined the effects of density fluctuation beyond the
simulation box comparing two box sizes of 50$h^{-1}$Mpc and
100$h^{-1}$Mpc and found that the effects of box size on lensing PDFs
are not significant.

First we have studied the convergence PDF. We have found that the mean
convergence $\langle \kappa \rangle$ measured in the source plane is
not zero, as often assumed in various analysis of cosmological lensing
effects, but systematically has negative values.
The mean convergence is
found to be correlated well with the variance of the convergence, which
follows $\langle \kappa \rangle\simeq -2 \langle \kappa^2 \rangle$ as
expected from simple consideration. 
Meanwhile the mean convergence $\langle\kappa\rangle$ measured in
the image plane is zero within small scatters less than 10$^{-3}$.
Therefore, we have to use appropriate statistics depending on the situation.
We have found that the modified
log-normal model of Das \& Ostriker (2006) reproduces our simulation
results quite well, except at the high convergence tail where the model
slightly underestimate the PDF. 

Next we have shown how the magnification PDFs are closely related with
the convergence PDFs. Specifically, we have pointed out that the simple
relation $\mu=(1-\kappa)^{-2}$, which approximates the correlation
between convergence and magnification seen in our simulations, can be
used to convert the convergence PDF to magnification PDF (or vice
versa). Surprisingly, the magnification PDF obtained via this conversion
agrees very well with simulation results up to very high-magnification
tail of $\mu\sim 10$. In light of this finding, we have presented a
simple analytic model of the magnification PDF based on the convergence
PDF of Das \& Ostriker (2006). We have also presented a simple fitting
formula of the shear PDF, which is shown to reproduce the simulation
results very well.  As explicit applications of these PDFs, in this
paper we have discussed strong lensing probabilities and made
quantitative predictions for the magnification effects on observed
luminosity functions for distant sources.

To summarize, the convergence PDF is computed by Equations
(\ref{lognormal_Das}), (\ref{Das_params}), and (\ref{kappa_empty}), the
magnification PDF is computed by Equation (\ref{Das_Oguri_eq}) via the
convergence PDF, and the shear PDF is computed by Equations
(\ref{shearPDF_oguri}) and (\ref{shearPDF_oguri_param}).  All these
models require the variances of the convergence and shear.  Under the
weak lensing approximation, we can compute them analytically as \beqa
\langle \kappa^2 \rangle = \langle \gamma^2 \rangle = \frac{9}{8 \pi}
H_0^4 \Omega_m^2 \int_0^{z_s} \frac{dz}{H(z)} \left( 1+z \right)^2
\nonumber \\ \times \left[ \frac{r(z) r(z,z_s)}{r(z_s)} \right]^2 \int
dk k P(k,z) \left[W(kr(z)\theta)\right]^2, \eeqa where $W(x)$ is the
window function corresponding to the shape and size of the source or the
smoothing scale of interest, e.g., $W(x)=2J_1(x)/x$ for the top-hat and
$W(x)=\exp(-x^2/2)$ for the Gaussian window function.

Our numerical simulations consider dark matter particles only, while
the baryonic effects become important in the galactic scale. For
instance, Hilbert et al. (2007) investigated the effects of stellar
masses in dark halos on the ray-tracing simulation through the
Millennium simulation. They used semi-analytical galaxy models from
the halo merging history  to investigate the baryonic effects, and
showed that the baryon increases the magnification PDF by a few ten
percents at $\mu=100$ for the source redshift of $z_s=2.1$, 
and enhance the strong lensing probability by an order of magnitude at
$z_s>1$. We are planning to include the baryonic components by simply
placing galaxies (bulge and disk components) in dark halos. Such
simulations should also allow us to study correlations between the
magnification map and the distribution of foreground objects such as
galaxies and clusters (Takahashi et al. in preparation).
Similarly, the dark matter substructures would affect
 the small-scale power spectrum and the weak lensing observables.
In fact,  Hangan et al. (2005) showed that the subhaloes
 enhance the power spectrum at $k>100h(\mbox{Mpc})^{-1}$.
Hence, the unresolved subhaloes in our simulation
 could affect our prediction.

It is known that the statistical property of the random-Gaussian field
is fully characterized by the power spectrum. However, the
characterization of the non-Gaussian fields, such as the lensing field
studied in this paper, require information on higher-order correlation
in addition to the power spectrum. In fact, it has been shown that the
power spectrum contains little information in non-Gaussian regime
(e.g., Rimes \& Hamilton 2005). Therefore, the PDFs studied in this
paper are expected to contain useful additional information on 
the statistical property of the lensing map.
Recently, Neyrinck et al. (2009) showed that gaussianizing the
one-point distribution function of the matter density fluctuation by
using a modified log-normal transformation, $\delta \rightarrow
\ln(1+\delta)$, increase the signal-to-noise ratio (S/N) of the
transformed power spectrum. The result implies that such
gaussianization provides a means of extracting non-Gaussian
information and recovering the information content of the power
spectrum. Similarly, Seo et al. (2011) showed that the cosmological
information in the convergence power spectrum is recovered by using a
modified logarithmic transform of the convergence field.
The accurate PDFs presented in the paper might be useful in this
regard.

Our simulation results, including two-dimensional maps of the
convergence, shear, magnification, and PDFs of these quantities, 
are publicly available at 
http://cosmo.phys.hirosaki-u.ac.jp/takahasi/raytracing/.

\acknowledgments

We greatly appreciate Takahiro Nishimichi for kindly providing
parallelised 2nd-order Lagrangian perturbation theory code.
M.O. thanks Hirosaki University for its warm hospitality during his
visit, where this work was initiated.
M.S. is supported by Grants-in-Aid for Japan Society for the Promotion of
Science (JSPS) Fellows.
This work was supported in part by Grant-in-Aid for Scientific Research
on Priority Areas No. 467 ``Probing the Dark Energy through an
Extremely Wide and Deep Survey with Subaru Telescope'',
 by the Grand-in-Aid for the Global COE Program
 ``Quest for Fundamental Principles in the Universe: from Particles
 to the Solar System and the Cosmos'' from the Ministry of Education,
 Culture, Sports, Science and Technology (MEXT) of Japan,
 by the MEXT Grant-in-Aid for Scientific Research on Innovative Areas
 (No. 21111006),
by the FIRST program "Subaru Measurements of Images and Redshifts
(SuMIRe)", World Premier International Research Center Initiative (WPI
Initiative) from MEXT of Japan, and by Grant-in-Aid for Scientific Research
from the JSPS (23740161).
Numerical computations were carried out on Cray XT4 at Center for
 Computational Astrophysics, CfCA, of National Astronomical Observatory
 of Japan.

\appendix

\section{Box size effect}

\begin{figure*}
\epsscale{1.1}
\plotone{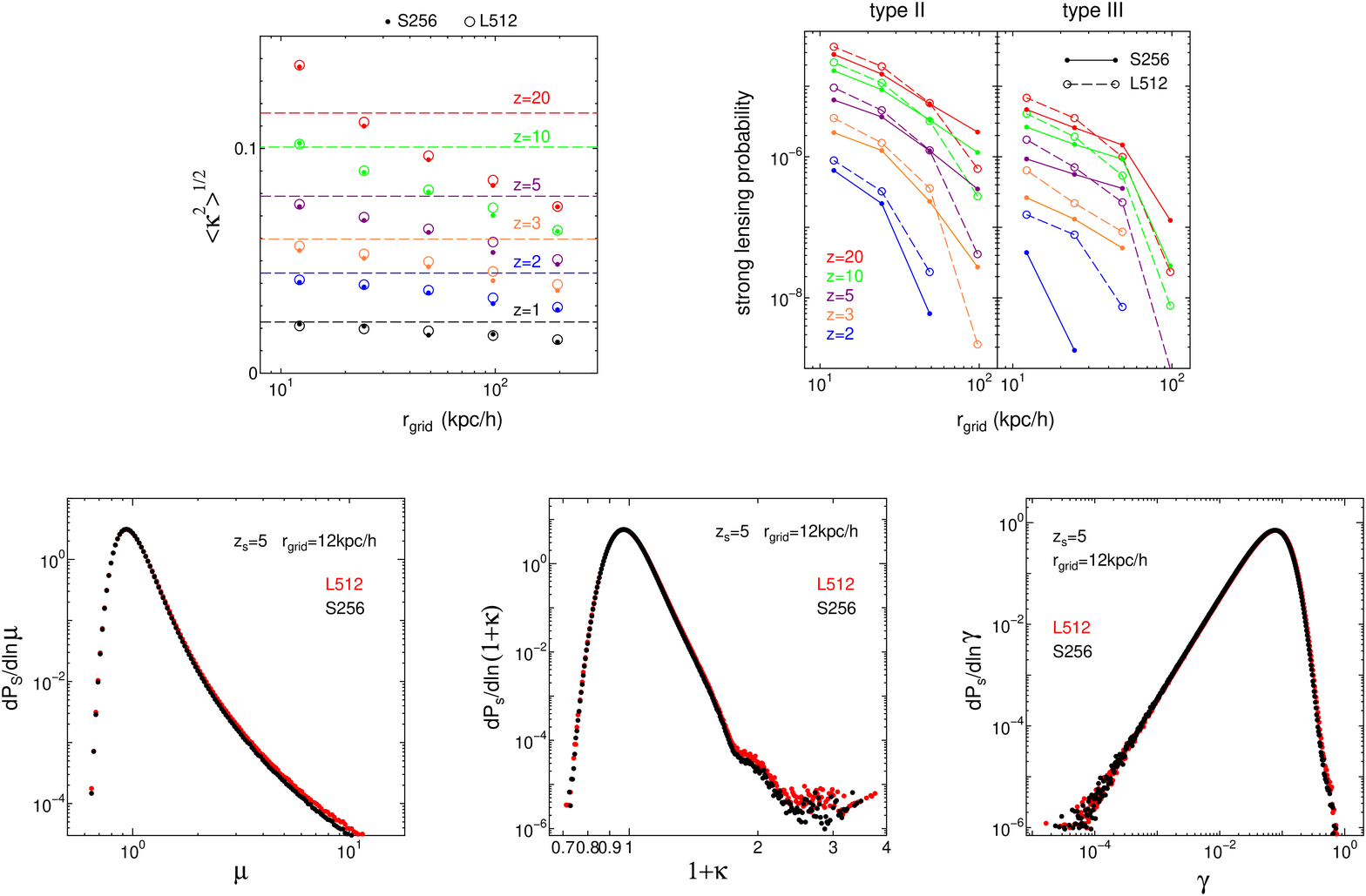}
\caption{
 Comparison of two simulations with a different box size, the small
 box $L=50h^{-1}$Mpc (model S256) and the large box $L=100h^{-1}$Mpc
 (model L512). These two simulations have the same mass and force
 resolutions (see Table~\ref{table1}). Top left panel shows the root
 mean square of the convergence, the dots are the results for S256 and
 the circles are for L512. Top right panel plots the strong lensing
 probability. Bottom panels display the magnification PDF (bottom left),
 the convergence PDF (bottom middle), and the shear PDF (bottom right).
 The black symbols are for S256, while the red symbols are for L512. 
 These result indicate that the box size effect is negligible.
}
\label{fig_boxsize}
\vspace*{0.5cm}
\end{figure*}

Throughout this paper, we have shown the results for the simulation box
 $50h^{-1}$ Mpc on a side (Model S256, S512 and S1024, see Table~\ref{table1}). 
However, the density fluctuation larger than the box size may affect
our results. In order to check such potential box size effect, we
compare our simulation results with those of the larger simulation box
of $100h^{-1}$ Mpc (Model L512). We compare the results for the model
S256 (smaller box) and L512 (larger box) in Figure~\ref{fig_boxsize}.
Note that the two models have the same mass and force resolutions.
We check the root mean square of the convergence (top left), 
strong lensing probability (top right), and PDFs of magnification
 (bottom left), convergence (bottom middle) and shear (bottom right). 
We find that two results agree well with each other. 
A possible exception is the strong lensing probability and PDFs 
at the high magnification and convergence tail, which appear to be
slightly enhanced in the larger box size simulation.
The differences is presumably because more massive halos are formed
for larger box simulations. In either case, our results here confirm
that the box size effect is insignificant. 

\if{}

\section{Updating the halo-fit model}

To calculate the variance of the convergence $\langle \kappa^2 \rangle$
 analytically in Equation~(\ref{rms_kappa}), we need the accurate nonlinear
 power spectrum.
Although the halo-fit model by Smith et al. (2003) is frequently used to
 calculate the nonlinear power spectrum,
 the several authors already pointed out that this formula
 does not agree with the results of recent high-resolution simulations.
Vale \& White (2004) suggested that the halo fit predicted the smaller
 power than their numerical results for small scale fluctuations
 (see also Sato et al. 2009).
Heitmann et al. (2010) ran high-resolution simulations, so called the
 ``Coyote Universe'' simulation, and showed that the halo fit is $\sim 5 \%$
 smaller than their numerical results at $k \sim 1h$Mpc$^{-1}$ and the
 deviation increases further for larger $k$. 
Lawrence et al. (2010) provided an emulator to calculate the nonlinear
 power spectra for arbitrary cosmological parameters.
The emulator interpolates the nonlinear power spectra of the
 ``Coyote Universe'' simulation for some different cosmological models.
However their emulator can be used only for small $k (\le 1h{\rm Mpc}^{-1})$
 and low redshift $0 \le z \le 1$, so it is not useful for weak lensing
 analysis such as cosmic shear.

Hence, in this section, we re-calculate the best-fit parameters in the
 halo-fit model using our numerical results.
We use the power spectra in model S512 and L512 for $z \le 10$.
There are $138$ realizations in S512 and $69$ realizations in L512
 for $z \le 10$, so we have $207$ independent realizations in total. 
There are $27$ parameters in the halo fit and we determine these parameters
 using the standard $\chi^2$ analysis:
\beq
  \chi^2 (\bfx) = \sum_z \sum_{k_1,k_2} {\rm{cov}}^{-1}(k_1,k_2,z) \left[
    P_{\rm sim}(k_1,z) - P_{\rm hf}(k_1,z;\bfx) \right] \left[
    P_{\rm sim}(k_2,z) - P_{\rm hf}(k_2,z;\bfx) \right],
\eeq 
where $\bfx=(x_0,\cdot \cdot \cdot,x_{26})$ are the fitting parameters,
 $P_{\rm sim}$ is our simulation results, $P_{\rm hf}$ is the halo-fit
 model.
The covariance matrix of the power spectrum is given by (e.g. Scoccimarro
 et al. 1999; Meiksin \& White 1999; Takahashi et al. 2009),
\beq
  {\rm{cov}}(k_1,k_2,z) = \frac{2}{N_k} P_{\rm sim}^2 (k_1,z)
 \delta_{k_1,k_2} + \frac{1}{V} T(k_1,k_2,z),
\eeq
where $N_k$ is the number of Fourier modes, $\delta_{k_1,k_2}$ is the
 Kronecker-type delta function, $V$ is the volume of the simulation box,
 and $T$ is the trispectrum.
The number of modes in the binwidth $\Delta k$ is
 $N_k=V k^2 \Delta k/(2 \pi^2)$.
The first term represents the Gaussian error contribution, and this term
 has only diagonal elements because the different wavenumbers are
 uncorrelated for the Gaussian density fluctuation (e.g. Feldman et al.
 1994).
The relative error of $P(k)$ is simply given by
 the inverse of the square root of the number of Fourier modes $N_k$.
The second term represents the non-Gaussian errors that include
 correlations between power spectra at different wavenumbers arising
 from non-linear mode coupling during gravitational evolution
 (e.g. Scoccimarro et al. 1999; Meiksin \& White 1999). 
To calculate the trispectrum, we use the halo model (Cooray \& Hu 2001).
The trispectrum consists of four terms, from 1- to 4-halo terms.
Here we consider only 1-halo term because for small scale
 $k>1h$Mpc$^{-1}$ the 1-halo term dominates (Cooray \& Hu 2001).

The best fit parameters are given by,
\beqa
  && \log_{10} a_n=1.4861+1.8369n+1.6762n^2+0.7940n^3+0.1670n^4-0.6206C \\
  && \log_{10} b_n=0.9463+0.9466n+0.3084n^2-0.9400C \\
  && \log_{10} c_n=-0.2807+0.6669n+0.3214n^2-0.0793C \\
  && \gamma_n=0.8649+0.2989n+0.1631C \\
  && \alpha_n=1.3884+0.3700n-0.1452n^2 \\
  && \beta_n=0.8291+0.9854n+0.3401n^2  \\
  && \log_{10} \mu_n=-3.5442+0.1908n \\
  && \log_{10} \nu_n=0.9589+1.2857n
\eeqa

where values in the parentheses are the parameters in Smith et al. (2003).
$n$ is the effective spectrum index and $C$ is its curvature (see Appendix C
 in Smith et al. (2003) for definition of their parameters).

it is easy to replace the above parameters.

Figure  shows the power spectra at $z=0,1,3,10$.

\fi{}

\clearpage

\end{document}